# Sonochemically Boosted Hydrogen Evolution Activity of Janus TMD Monolayers

*Rayantan Sadhukhan[1†], Md Tarik Hossain[2†], Julian Picker[2], Mahdi Ghorbani-Asl[3], Christof Neumann[2], Arkady V. Krasheninnikov[3], Tharangattu N. Narayanan[1]*, Andrey Turchanin[2,4]**

[1]Tata Institute of Fundamental Research Hyderabad,
Serilingampally Mandal, Hyderabad, Telangana 500046, India.

[2]Friedrich Schiller University Jena, Institute of Physical Chemistry,
Lessingstr. 10, 07743 Jena, Germany.

[3]Helmholtz-Zentrum Dresden-Rossendorf, Institute of Ion Beam Physics and Materials Research,
Bautzner Landstr. 400, 01328 Dresden, Germany.

[4]Center for Energy and Environmental Chemistry, Philosophenweg 7a, 07743 Jena, Germany.

[†] = *Contributed equally*

Present address for J.P.: MAX IV Laboratory, Lund University, Fotongatan 2, 224 84 Lund

**Corresponding Authors**:

Tharangattu N. Narayanan (tnn@tifrh.res.in)

Andrey Turchanin (andrey.turchanin@uni-jena.de)

**Abstract**

2D electrocatalysts that enable hydrogen evolution at low overpotentials offer an attractive alternative to expensive platinum-based systems. Here, we report the growth of Janus transition metal dichalcogenide (TMD) monolayers (MLs), SeMoS and SeWS, on Au foils using chemical vapor deposition, and systematically compare their catalytic properties in the context of hydrogen evolution reaction (HER) with those of their parent TMDs. The Janus MLs exhibited significantly enhanced catalytic performance relative to the parent TMDs. Furthermore, these MLs on Au foils were subjected to sonochemical treatment in polar and non-polar solvents, in which the treatment with polar solvents led to a substantial improvement in the HER activity of Janus MLs. In particular, SeMoS MLs treated with water showed a low overpotential of ~63 mV, a Tafel slope of ~42 mV/dec, and an exchange current density of ~$10^{-3}$ mA $cm^{-2}$, approaching that of platinum. Analyses indicate that enhanced electrocatalytic activity is associated with tensile strain induced by Au surface restructuring and the formation of defects in Janus MLs, as shown by experimental observations and by density functional theory calculations. The enhancement in catalytic performance due to sonochemical treatment emphasizes the importance of our results for developing novel catalytic systems for HER based on Janus 2D materials.

**KEYWORDS**

Janus TMD; 2D electrocatalysts; chemical vapor deposition; hydrogen evolution reaction; DFT calculations. sonochemistry

## Introduction

Hydrogen ($H_2$) is considered to be one of the most valuable industrial chemicals.[1-4] Platinum and other platinum group metals are the current benchmarks for hydrogen generation *via* water electrolysis;[5] however, their high cost poses a significant barrier to large-scale, low-cost implementation.[6] Transition metal dichalcogenides (TMDs) have emerged as a promising class of materials, offering high thermodynamic efficacy for the electrocatalytic hydrogen evolution reaction (HER) at lower cost.[7-9] $MoS_2$ and $MoSe_2$ have demonstrated near-optimal hydrogen adsorption free energy ($\Delta G_H$), with values of approximately +0.08 eV[10] and -0.04 eV[11], respectively, positioning them as attractive alternatives to platinum-based catalysts ($\Delta G_H \approx 0.03$ eV)[12]. However, kinetically - decided by the number of active catalytic centers per area (deduced from the exchange current density), the 2H-phase TMD monolayers (MLs) are not comparable to Pt due to the low number of active sites for HER. Typically, their HER activity is limited to their edges and point defects, leaving the basal plane largely inert.[13, 14] The primary strategies to overcome this include defects and edge engineering[15-17], phase engineering[18, 19], creating chalcogen vacancies and applying tensile strain[20]. Specifically, chemical vapor deposition (CVD) on substrates featuring Au nanocones has been used to induce capillarity-force-driven tensile strain, to achieve the optimum value of $\Delta G_H$ ( $\approx 0.08$ eV) in the basal plane of 2H phase $MoS_2$ MLs with S vacancies.[20] Nevertheless, to fully realize the potential of 2D TMD catalysts, there remains a need to develop efficient materials and simple methods for tuning their catalytic properties.

Recent advances in the engineering of ML TMDs have enabled the development of Janus materials, in which one of the chalcogenide layers is replaced by another chalcogen (e.g., SeMoS).[21] The disparity in the electronegativity between the chalcogen elements in the top and bottom layers creates an asymmetric dipole distribution, generating a "colossal vertical electric field" within the Janus ML.[22] These Janus MLs are predicted to reach thermoneutrality with high kinetic activity, with an exchange current density of the order of $10^{-4}$ mA $cm^{-2}$, compared to $10^{-5}$ mA $cm^{-2}$ for conventional ML $MoS_2$.[23] In addition, different chalcogen layer exposure results in distinct catalytic properties. In this context, Janus SeMoS MLs with sulfur as the bottom layer adjacent to the $SiO_2$/Si substrate and selenium as the top layer exposed to the electrolyte exhibit better catalytic HER activity due to the change in the proton adsorption energy on the respective chalcogens, induced *via* the dipole orientation[24]. Despite these advances, challenges in their scalable synthesis and tunability of catalytic properties remain elusive.

Here, we report the synthesis and demonstrate the electrocatalytic potential of 2H phase SeMoS and SeWS Janus MLs (Se as the top layer) grown on Au foil, building on a thermodynamic-equilibrium-driven synthesis approach[25]. The HER catalytic activity was assessed in an acidic medium (0.5 M $H_2SO_4$, pH ~ 0), where it was systematically compared with conventional TMD MLs, such as $MoS_2$, $MoSe_2$, $WS_2$, and $WSe_2$, directly grown on Au foil. To further enhance the innate HER efficacy of Janus MLs, we introduce a sonochemical-activation treatment, involving ultrasonication of the as-grown Janus MLs on Au foil in polar solvents. Combined experimental observations and density functional theory (DFT) calculations indicate that defects and tensile strain in Janus materials underpin the observed improvement in HER activity, offering an effective method to tune the catalytic properties of Janus TMD MLs.

**Results and Discussion**

Janus TMD MLs of SeMoS and SeWS were directly grown on Au foil using chemical vapour deposition (CVD).[25] The synthesis involves a one-pot two-step CVD process. In the first step, MLs of $MoSe_2$ and $WSe_2$ are grown on Au foils. In the second step, the bottom selenium layers are replaced by sulfur through a thermodynamic-equilibrium-driven exchange[25], resulting in the formation of SeMoS and SeWS Janus MLs. Detailed growth method and temperature-time profiles are provided in the Supporting Information and in Supplementary **Fig. S1**.

Next, the synthesized SeMoS and SeWS Janus materials, **Figs 1a** and **1b,** were characterized using optical microscopy, Raman spectroscopy, X-ray photoelectron spectroscopy (XPS), and scanning electron microscopy (SEM). Optical microscopy images (**Fig. 1c, d**) show Janus TMD crystals with lateral dimensions of approximately 10-30 μm. The Raman spectra display distinct $A_1^1$ modes at 291 $cm^{-1}$ for SeMoS (**Fig. 1e**) and 283 $cm^{-1}$ for SeWS (**Fig. 1f**), consistent with previous reports[25-27], thereby confirming the successful formation of 2H Janus MLs. The chemical composition and the asymmetric structure of SeWS are investigated by performing angle-resolved X-ray photoelectron spectroscopy (ARXPS) on as-grown samples on Au foils (Supplementary **Fig. S2**). The binding energies and intensities of W 4f, S 2p, and Se 3d peaks confirm the formation of ML Janus SeWS (Supplementary **Fig. S2a-d**). ARXPS, exploiting enhanced surface sensitivity at higher photoelectron emission angles (θ), shows from the relative intensity (RI) variations of Se, W, S, and Au that Se resides in the top layer, while W and S lie beneath (see Supplementary **Fig. S2e**). Thus, the Se layer is the top layer in our Janus structures. The atomic ratios of Se:Mo:S and Se:W:S are estimated to be (1.0 ± 0.2):1.0:(1.9 ± 0.2) and (0.8 ± 0.2):1.0:(3.4 ± 0.2), indicating an excess of sulfur atoms, possibly due to chemisorption on the Au substrate.[25] For comparison, conventional $MoS_2$, $MoSe_2$, $WS_2$, and $WSe_2$ MLs were also grown on Au foils using the CVD method and characterized by optical microscopy and Raman spectroscopy, as depicted in Supplementary **Fig. S3**.

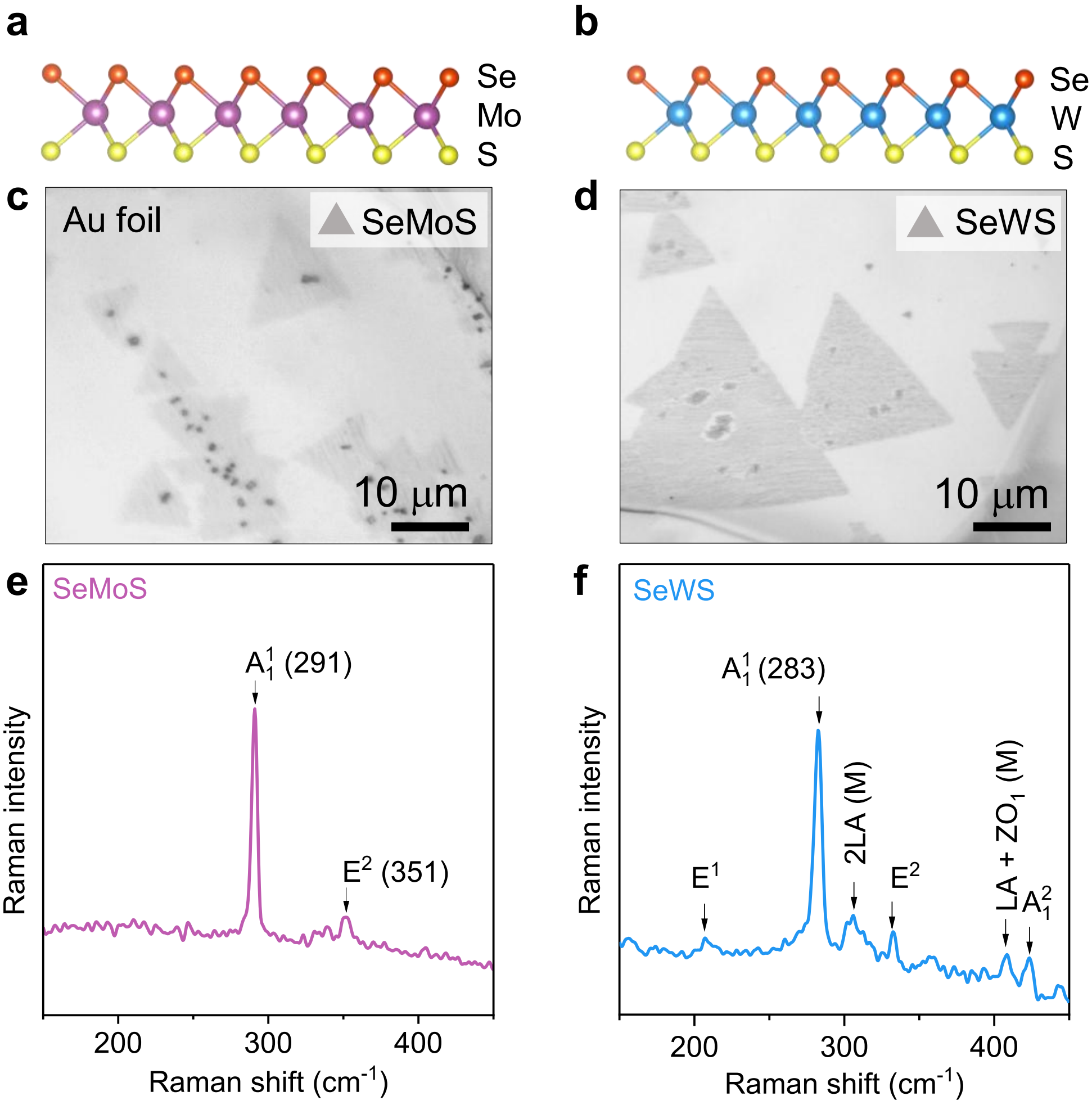

**Figure 1: Characterization of CVD-grown Janus SeMoS and SeWS MLs on Au foil.** Cross-sectional atomic model of (a) SeMoS. (b) SeWS. (c, d) Optical microscopy images corresponding to SeMoS and SeWS, respectively. (e, f) Raman spectra recorded at room temperature using 532 nm excitation wavelength on SeMoS and SeWS, respectively.

After the characterization, we systematically evaluated the HER catalytic activity of Janus TMDs in comparison with their conventional (non-Janus) counterparts, examining both as-grown samples and those subjected to various treatment protocols (**Figs. 2** and **3** for Mo- and W-based TMDs, respectively). The three different treatment protocols include solvent immersion, immersion followed by heating (90 °C), and sonochemical treatment (40 kHz, 30 min) in solvents spanning a wide polarity range (more details are in the supplementary information, page S3). The electrocatalytic properties were studied using a three-electrode experimental setup (refer to Supplementary **Fig. S4**). HER performance was quantified in 0.5 M $H_2SO_4$ using the overpotential required to achieve a current density of 10 mA $cm^{-2}$, denoted as $\eta_{10}$ and determined from the linear sweep voltammetry (LSV) measurements. As-grown $MoS_2$ and $MoSe_2$ show $\eta_{10}$ values of ~ 452 mV and ~ 480 mV, respectively, whereas the as-grown Janus SeMoS exhibits a significantly lower $\eta_{10}$ value of 232 mV (**Fig. 2a**), consistent with the higher hydrogen adsorption efficiency as a result of the intrinsic dipole moment in the Janus materials[23]. Solvent immersion, as well as combined immersion and heat treatment, induces sluggish changes in the HER catalytic performance of both Janus and conventional TMDs (see Supplementary **Figs. S5, S6,** and **S7)**.

Interestingly, sonochemical treatment in $H_2O$ significantly improves the HER catalytic activity of Janus SeMoS MLs (SeMoS ($H_2O$) in **Fig. 2a**), reducing the $\eta_{10}$ value to ~ 63 mV. In contrast, sonochemically $H_2O$-treated $MoS_2$ and $MoSe_2$ show only modest reductions (**Fig. 2a)** in $\eta_{10}$ to 408 mV and 453 mV, respectively, showing minimal enhancement relative to the huge improvement observed in Janus MLs. Control experiments show that bare Au foil and sonochemically $H_2O$-treated Au foil exhibited high $\eta_{10}$ values (576 mV and 584 mV, respectively) (**Fig. 2a**). To further understand the role of solvent polarity, both polar and non-polar solvents (Supplementary **Table S1**), including benzene ($C_6H_6$), chloroform ($CHCl_3$), isopropanol (IPA, $C_3H_8O$), and $H_2O$, were employed for sonochemical treatment (**Fig. 2b**).

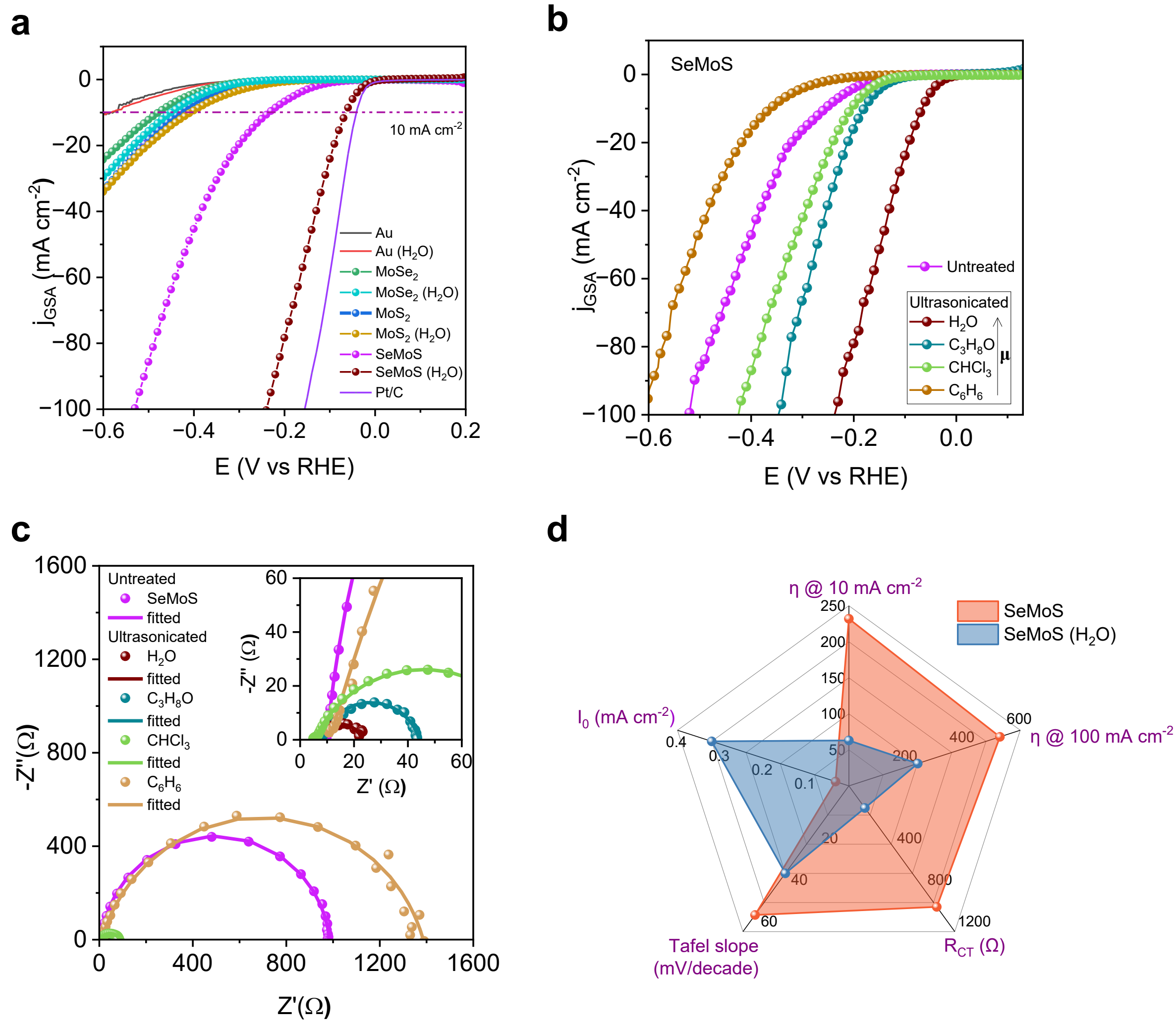

**Figure 2: HER polarization curves for different Mo-based TMDs and their respective impedance spectroscopies.** (a) Linear sweep voltammetry (LSV) curves of different Mo-based TMD MLs before and after sonication in water ($H_2O$), showing changes in HER activity. (b) HER LSV curves after sonication in $H_2O$ (very polar), isopropanol (IPA, $C_3H_8O$, less polar), chloroform ($CHCl_3$, even less polar), or benzene ($C_6H_6$, non-polar) for Janus SeMoS. 'μ' is the relative dipole moment of the solvents. (c) Electrochemical impedance spectroscopy (EIS) data for Janus SeMoS after sonication in different solvents. The inset picture is the magnified version of the fitted EIS plots showing very small charge transfer resistance for the catalysis activity. (d) Performance radar chart comparing the catalytic activity of SeMoS and SeMoS ($H_2O$). SeMoS ($H_2O$) has lower overpotential (η), lower Tafel slope (faster reaction kinetics), lower charge transfer resistance ($R_{CT}$), and higher exchange current density ($I_0$) than SeMoS electrocatalyst.

Among these, $H_2O$ yields the most significant enhancement, bringing the $\eta_{10}$ of SeMoS close to that of Pt (**Fig. 2a**). IPA and chloroform also improve HER activity, with $\eta_{10}$ values of ~157 mV and ~199 mV, respectively, consistent with their intermediate polarity (**Fig. 2b**). In contrast, non-polar benzene produces no improvement (**Fig. 2b**), with $\eta_{10}$ remaining high (~246 mV).

To further verify the faster HER process of sonochemically treated Janus layers, electrochemical impedance spectroscopy (EIS) studies were conducted. Details of the EIS are given in the Supporting Information (page S5) and in Supplementary **Fig. S8**. Noticeably, the charge transfer resistance ($R_{CT}$) for as-grown SeMoS is ~ 962 Ω (**Fig. 2c**), while that of SeMoS ($H_2O$) is ~12 Ω, indicating a gigantic reduction in the charge transfer resistance leading to an enhanced HER kinetics. The $R_{CT}$ values of SeMoS ($C_3H_8O$) and SeMoS ($CHCl_3$) were also reduced to ~ 34 Ω and ~ 86 Ω, respectively, consistent with the trends observed in the earlier LSV measurements. In contrast, the $R_{CT}$ for SeMoS ($C_6H_6$) is ~ 1155 Ω, indicating sluggish kinetics with the non-polar benzene-assisted sonochemical process. Thus, a clear correlation between solvent polarity and catalytic performance is observed (Supplementary **Fig. S10a**), with higher solvent polarity leading to lower overpotentials and reduced charge transfer resistance, thereby highlighting the critical role of solvent-mediated sonochemical effects in activating the catalytic properties of Janus TMDs.

The Tafel slopes for as-grown SeMoS and solvent sonicated SeMoS samples are presented in Supplementary **Fig. S10b**, where smaller slopes indicate higher catalytic activity and faster reaction kinetics. **Fig. 2d** compares the catalytic activity parameters, highlighting that the SeMoS ($H_2O$) catalyst exhibits markedly faster HER kinetics. This catalyst shows a Tafel slope of 42 mV $dec^{-1}$ as compared to 82 mV $decade^{-1}$ for as-grown Janus SeMoS and closer to the 38 mV $dec^{-1}$ observed for the Pt/C catalyst. Alongside, there is a 10-fold increase in the exchange current density ($I_0$), which brings SeMoS Janus $I_0$ value of ~$10^{-3}$ mA cm$^{-2}$, similar to that of Pt.

A similar trend is observed for SeWS MLs. As-grown SeWS exhibits $\eta_{10} \approx 259$ mV, which decreases to ~ 92 mV after sonication in $H_2O$ (**Fig. 3a**), along with a significant reduction in $R_{CT}$ (~ 1664 → ~ 25 Ω) (**Fig. 3c**), a lower Tafel slope (~ 84 → ~ 51 mV dec$^{-1}$) (**Fig. S11**), and a 10-fold enhancement in exchange current density (**Fig. 3d**). Intermediate activity is obtained in IPA and chloroform ($\eta_{10} \approx$ 166 and 201 mV; $R_{CT} \approx$ 77 and 135 Ω), while benzene results in sluggish performance ($\eta_{10} \approx 388$ mV; $R_{CT} \approx 2553$ Ω) (**Figs. 3b and 3c**). Both $\eta_{10}$ and $R_{CT}$ change systematically with solvent polarity. Intermediate solvent sonication (IPA and chloroform) again yields moderate improvements, whereas benzene results in sluggish performances, showing the same polarity-driven activation observed in SeMoS but with reduced overall activity. In contrast, $WS_2$ and $WSe_2$ exhibit negligible change in HER to identical treatments (**Figs. 3a** and **S7b)**, highlighting the unique sensitivity of Janus structures to sonochemical activation.

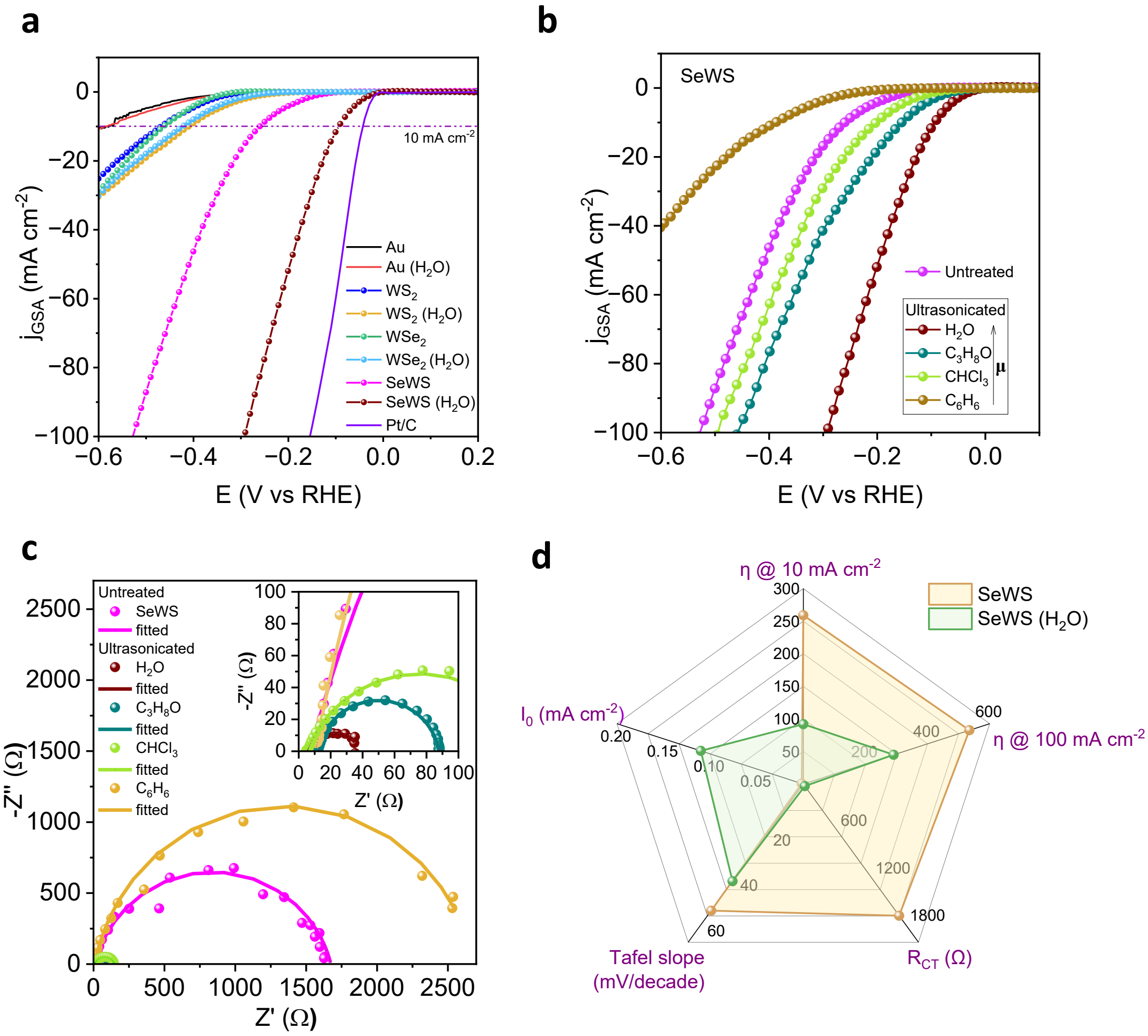

**Figure 3: HER polarization curves for different W-based TMDs and their respective impedance spectroscopies.** (a) Linear sweep voltammetry (LSV) curves for different W-based TMD MLs after sonication in water ($H_2O$), showing changes in their HER activity. (b) HER LSV curves after sonication in water (strongly polar), isopropanol (IPA, $C_3H_8O$, less polar), or benzene (non-polar) for SeWS Janus. 'μ' is the relative dipole moment of the solvents. (c) EIS data for Janus SeWS after sonication in different solvents. The inset picture is the magnified version of the fitted EIS plots showing very small charge transfer resistance for the catalysis activity. (d) Performance radar chart comparing the catalytic activity of SeWS and SeWS ($H_2O$). SeWS ($H_2O$) has lower overpotential (η), lower Tafel slope (faster reaction kinetics), lower charge transfer resistance ($R_{CT}$), and higher exchange current density ($I_0$) than the SeWS electrocatalyst.

To shed light on the underlying mechanism, we conducted XPS, Raman, and SEM analyses on as-grown Janus SeMoS MLs before and after the sonochemical treatment in mainly $H_2O$, where the maximum change in HER activity was observed. Polar solvents are known to interact strongly with Janus TMDs *via* their intrinsic dipole moments, sometimes causing delamination from weakly adhesive substrates such as $SiO_2$.[28] Here, possibly the direct growth on Au provides stronger adhesion between Au and the bottom S layer of Janus, preventing detachment despite solvent exposure, even after ultrasonication.[29] XPS results (Supplementary **Fig. S12**) show that all elements of the Janus SeMoS are present after sonication with no significant chemical shifts, indicating that the Janus structure remains chemically stable. Moreover, the Raman modes of both the treated Janus SeMoS and SeWS showed a systematic redshift in line with the increase in the polarity of the solvent used during sonication, as can be seen in **Fig. 4** and **Fig. S13**. For example, the $A_1^1$ peak (**Fig. 4a**), corresponding to the out-of-plane vibration, shifts from 289 $cm^{-1}$ to 286 $cm^{-1}$ and 287 $cm^{-1}$ for SeMoS ($H_2O$) and SeMoS ($C_3H_8O$), respectively, while no shift is observed for SeMoS ($C_6H_6$). The redshift is indicative of tensile strain[30], which may arise from the sonochemical treatment. A similar trend is observed for SeWS, where the $A_1^1$ mode redshifts from 284 $cm^{-1}$ to 281 $cm^{-1}$ upon treatment in $H_2O$ (**Fig. 4b**), suggesting introduction of tensile strain to the Janus MLs. SEM imaging further reveals increased surface texturing of both the Janus layers and the Au substrate after sonication (Supplementary **Fig. S14**). Besides that, a large number of nanometer-sized cracks and irregular pores are created in the Janus MLs by polar solvent treatment, whereas the sample treated with non-polar solvent remains unchanged (Supplementary **Fig. S15**). This way, polar solvent treatment creates more active sites on the basal plane of Janus MLs, which benefits HER catalytic activity. Notably, the intrinsic dipole moment of Janus MLs likely makes them more sensitive to interactions with polar solvents[28] than nonpolar ones. This highlights that ultrasonication in a polar solvent of Janus layers on Au foil is an effective way to form

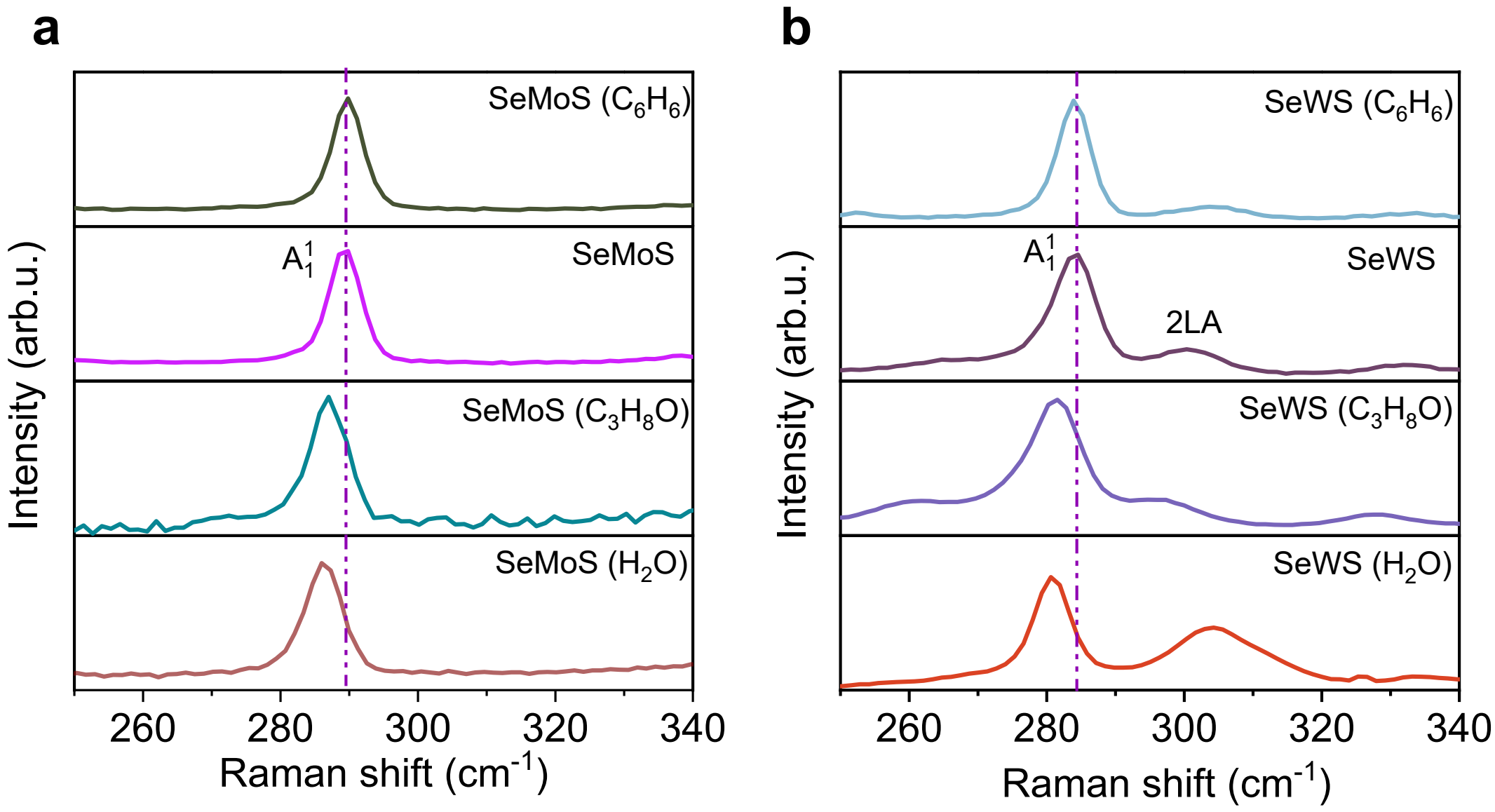

**Figure 4: Raman A′1 mode shifts in Janus SeMoS and SeWS before and after polar and non-polar solvent treatments.** (a) Raman spectra of SeMoS highlighting A′1 mode shifts induced by solvent treatment. (b) Raman spectra of SeWS showing corresponding A′1 mode changes. Dashed-dotted lines indicate peak positions for visual reference.

edge-enriched Janus MLs that would favour its electrochemical performance.

To further probe the impact of the Au foil during the sonochemical treatment, Au foils were identically treated sonochemically in polar ($H_2O$) and nonpolar ($C_6H_6$) solvents, and cyclic voltammograms (CVs) and thin film X-Ray Diffraction (XRD) were performed. CV of pristine Au foil majorly exhibits characteristic oxidation peaks corresponding to the (100), (110), and (111) surfaces, whereas treatment in $H_2O$ enhances the relative contribution of the (111) surface at the expense of (100) and (110) surfaces (Supplementary **Fig. S16**). This surface modification is absent in benzene-treated samples. The electrochemical areas of Au ($H_2O$) (2.87 $cm^2$) and Au ($C_6H_6$) (2.69 $cm^2$) are found to be similar, indicating that the electrochemical surface area remained unchanged. Further CV analysis with a reduced potential window (1.6 V vs RHE) shows a marked decrease in the $Au_2O_3$ reduction peak for Au ($H_2O$) relative to Au ($C_6H_6$), confirming a more exposed (111) surface in Au ($H_2O$) samples. Thin-film XRD corroborates these findings, with a reduced $I_{110}/I_{111}$ intensity ratio (~0.14 for Au ($H_2O$) vs ~0.3 for Au ($C_6H_6$), indicating preferential enhancement of the (111) surface in polar solvent treatment (Supplementary **Fig. S17**). Collectively, these observations indicate that polar solvent-mediated ultrasonication induces tensile strain. As shown next by DFT calculations, the strain induced by the surface restructuring along with the formation of defects in Janus MLs leads to improvement of the catalytic activity.

The energetics of hydrogen adsorption and the corresponding HER activity were further investigated using DFT calculations, with computational details provided in the Supplementary Information. **Fig. S18** and Table **S2** summarize the calculated Gibbs free energy of hydrogen adsorption ($\Delta G_H$) for different configurations of the Janus SeMoS ML. Since $\Delta G_H$ is a key descriptor for HER activity, values approaching zero indicate more favorable catalytic performance because hydrogen adsorption and desorption become thermodynamically balanced. The freestanding Janus SeMoS ML exhibits relatively large positive $\Delta G_H$ values of 2.10 eV, indicating weak hydrogen adsorption and therefore limited intrinsic HER activity. Our calculated values are in good agreement with previously reported theoretical results for freestanding Janus SeMoS MLs.[31] After introducing the Au(111) substrate, the hydrogen adsorption free energy is substantially reduced, demonstrating that the metallic support plays an important role in activating the Janus layer. This enhancement primarily arises from interfacial charge transfer and hybridization between the electronic states of the Au substrate and the SeMoS monolayer, thereby modifying the electronic structure near the Fermi level and enhancing the reactivity of the adsorption sites.[29]

Furthermore, applying tensile strain leads to an additional reduction in $\Delta G_H$, shifting the adsorption free energy closer to the thermoneutral value. The strain-induced enhancement can be attributed to modifications in the local atomic arrangement and redistribution of the electronic states around the active sites. In particular, tensile strain shifts the states associated with Mo orbitals and strengthens the interaction between the hydrogen orbital and the surface electronic states, thereby stabilizing hydrogen adsorption (**Figs. S19 a,b**). As a result, the strained supported structure exhibits significantly improved HER performance compared with the pristine freestanding system. We additionally examined the HER activity of a defective configuration containing a Se vacancy on the non-interfacial surface of the Janus layer. The vacancy-containing system shows a significant decrease in $\Delta G_H$, from 1.67 to −0.14 eV, bringing the adsorption free energy very close to the ideal thermoneutral condition and comparable to that of the Pt(111) catalyst. The improved activity originates from the formation of localized defect states near the Fermi level, which enhance orbital hybridization between the adsorbed hydrogen atom and neighbouring Mo-*d* states (**Fig. S19c).** In addition, the presence of the vacancy together tensile strain further amplifies this effect through additional charge redistribution and interfacial coupling (**Fig. S19 d**). Moreover, the catalytic activity of the Janus layer is highly sensitive to their edges.[32] Overall, these results demonstrate that Janus TMDs on Au foils, with strain and defects introduced by a post-sonication step with polar solvents, enable a substantial enhancement in the HER catalytic activity of these MLs.

Finally, the electrochemical stability of the sonochemically treated Janus MLs on Au foils was evaluated through HER tests. As shown in **Figs. 5a** and **5b**, chronoamperometry (CA) measurements conducted at a constant applied potential of -350 mV over seven days demonstrate stable performance. The SeMoS ($H_2O$) sample maintained a current density of 235 mA $cm^{-2}$ at 350 mV, while SeWS ($H_2O$) exhibited a stable current density of 150 mA $cm^{-2}$ at 350 mV, confirming the electrochemical stability of these catalysts. Additionally, the water-treated (SeMoS ($H_2O$), SeWS ($H_2O$)) samples remained electrochemically stable after being subjected to heating at 90 °C for 12 hours, as shown in **Figs. 5c** and **5d**.

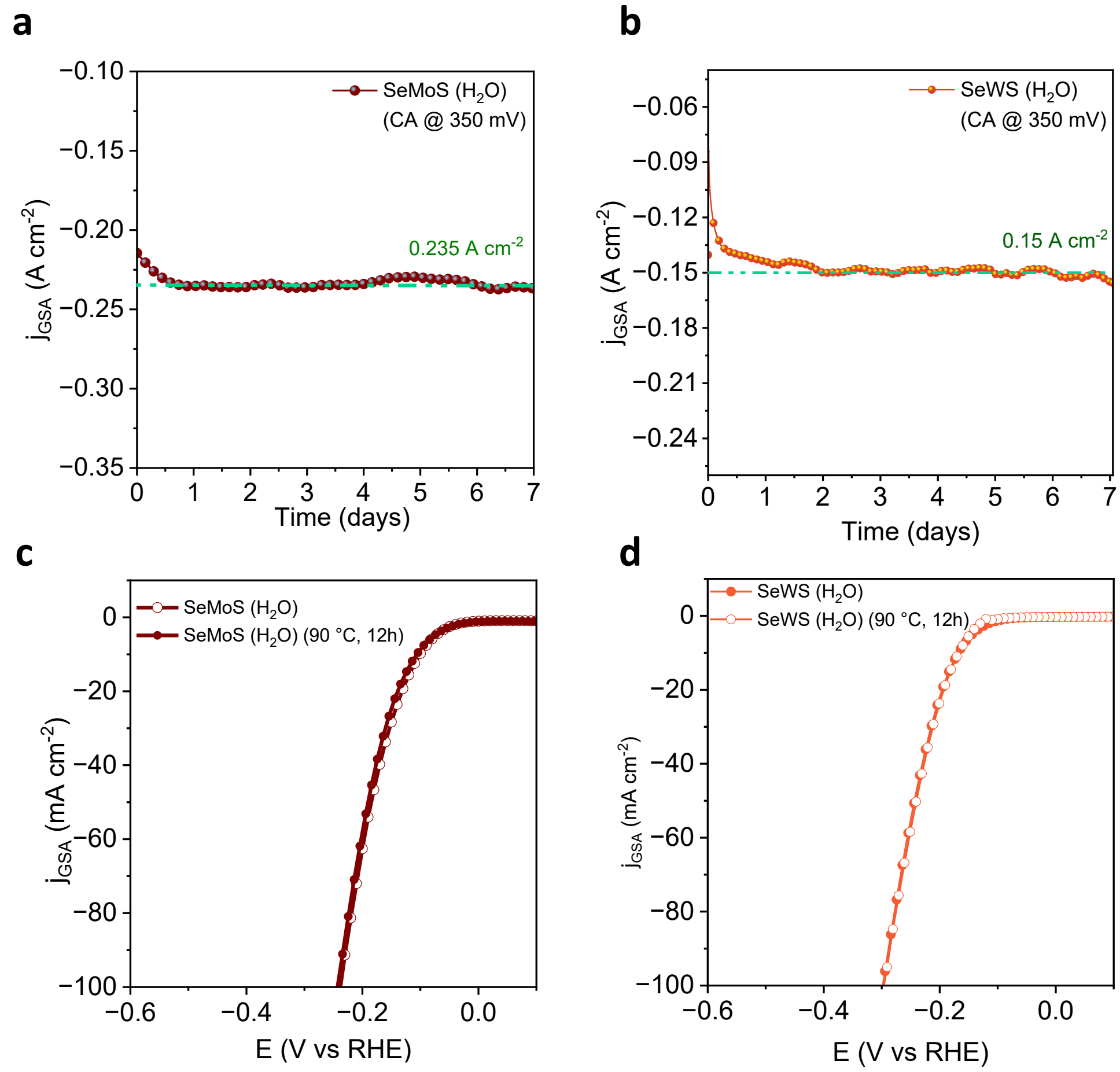

**Figure 5: Electrochemical and thermal stability of Janus TMDs.** Chronoamperometric measurements for long-term (seven days) electrolysis of SeMoS ($H_2O$) (a) and SeWS ($H_2O$) (b), and thermal stability of SeMoS (c) and SeWS (d) after heating at 90 °C for 12 hours.

In conclusion, we demonstrate a simple and scalable sonochemical approach to tune the HER catalytic activity of Janus TMD MLs directly synthesized on Au foils using CVD method. Ultrasonication in solvents of varying polarity enables systematic modulation of catalytic performance, with polar solvents - particularly water - yielding the most significant improvement. Notably, SeMoS MLs treated with $H_2O$ exhibit a low Tafel slope of 42 mV $dec^{-1}$, an exchange current density of $\sim10^{-3}$ mA $cm^{-2}$, and a low charge-transfer resistance of ~12 Ω, approaching the performance of Pt catalysts. Similar trends are observed for SeWS MLs. In contrast, the non-polar solvent-treated Janus MLs did not show such an enhancement in HER. Mechanistically, sonication in polar solvents increases defect density and also induces tensile strain in the Janus layers, thereby enhancing catalytic activity. DFT calculations further support the conclusion that strained Janus MLs with Se vacancies on the Au substrate significantly lower the HER energy barrier. Overall, our results establish sonochemical treatment as a simple and effective strategy for tuning the HER catalytic properties of Janus TMD MLs.

## Acknowledgements

M.T.H., J.P., C.N. and A.T. acknowledge the financial support of this work via the European Union Graphene Flagship project 2D Materials of Future 2DSPIN-TECH (No. 101135853), by the Deutsche Forschungsgemeinschaft (DFG, German Research Foundation) - CRC/SFB 1375 NOA "Nonlinear Optics down to Atomic scales" (Project B2, number 398816777), by BMBF Project SINNER Grant No. 16KIS1792, SPP2244 "2DMP" (Project TU149/21-1, 535253440), and DFG individual grant TU149/16-1 (464283495). R.S. thanks CataLight for providing the fellowship with which the visit at FSU Jena was possible for performing the experiments. T.N.N. acknowledges the financial support from the Alexander von Humboldt experienced researchers' fellowship which enabled him to work at FSU, Jena, Germany. T.N.N. and R.S. also acknowledge the support of the Department of Atomic Energy, India, under Project Identification No. RTI 4007. A.K. acknowledge DFG funding, projects KR 4866/11-1 and JO 1972/3-1 (Project No. 545786492), as well as the collaborative research center "Chemistry of Synthetic 2D Materials" CRC-1415-417590517. Generous CPU time grants from the Paderborn Center for Parallel Computing (PC2, Noctua 2 cluster, hpc-prf-def2dhet) and Gauss Centre for Supercomputing e.V. (http://www.gauss-centre.eu) at Höchstleistungsrechenzentrum Stuttgart, are greatly appreciated. We acknowledge Stephanie Höppener and Ulrich S. Schubert for providing access to the scanning electron microscope. The SEM facilities of the Jena Center for Soft Matter (JCSM) were established with a grant from the DFG.

## Author Contributions

T.N.N. and A.T. designed and directed the research. R.S. and M.T.H. performed the material synthesis, HER measurements, and contributed to the concept. Spectroscopy and microscopy measurements were conducted by R.S., M.T.H., J.P., C.N. and analyzed together with T.N.N.

and A.T. M.G.A. and A.V.K. conducted DFT calculations and their analysis. R.S., M.T.H., T.N.N. and A.T. wrote the manuscript with contributions of all co-authors.

## References

[1] N. Monnerie, P. G. Gan, M. Roeb, C. Sattler, *Int. J. Hydrog. Energy* **2020**, *45*, 26117.
[2] W. Liu, H. Zuo, J. Wang, Q. Xue, B. Ren, F. Yang, *Int. J. Hydrog. Energy* **2021**, *46*, 10548.
[3] M. Bacatelo, F. Capucha, P. Ferrão, F. Margarido, J. Bordado, *Int. J. Hydrog. Energy* **2024**, *63*, 382.
[4] M. Younas, S. Shafique, A. Hafeez, F. Javed, F. Rehman, *Fuel* **2022**, *316*, 123317.
[5] J. N. Hansen, H. Prats, K. K. Toudahl, N. Mørch Secher, K. Chan, J. Kibsgaard, I. Chorkendorff, *ACS Energy Lett.* **2021**, *6*, 1175.
[6] D. M. F. Santos, C. A. C. Sequeira, D. Macciò, A. Saccone, J. L. Figueiredo, *Int. J. Hydrog. Energy* **2013**, *38*, 3137.
[7] C. Zhu, D. Gao, J. Ding, D. Chao, J. Wang, *Chem. Soc. Rev.* **2018**, *47*, 4332.
[8] P. Prabhu, V. Jose, J.-M. Lee, *Matter* **2020**, *2*, 526.
[9] A. B. Laursen, S. Kegnæs, S. Dahl, I. Chorkendorff, *Energy Environ. Sci.* **2012**, *5*, 5577.
[10] T. F. Jaramillo, K. P. Jørgensen, J. Bonde, J. H. Nielsen, S. Horch, I. Chorkendorff, *Science* **2007**, *317*, 100.
[11] C. Tsai, K. Chan, F. Abild-Pedersen, J. K. Nørskov, *Phys. Chem. Chem. Phys.* **2014**, *16*, 13156.
[12] J. K. Nørskov, T. Bligaard, A. Logadottir, J. R. Kitchin, J. G. Chen, S. Pandelov, U. Stimming, *J. Electrochem. Soc.* **2005**, *152*, J23.
[13] Y. Cao, *ACS Nano* **2021**, *15*, 11014.
[14] D. Voiry, J. Yang, M. Chhowalla, *Adv. Mater.* **2016**, *28*, 6197.
[15] Z. Gan, R. Sadhukhan, C. Neumann, N. Mohandas, E. Najafidehaghani, M. Mundszinger, J. Biskupek, U. Kaiser, T. N. Narayanan, A. George, A. Turchanin, *Mater. Today Catal.* **2024**, *6*, 100060.
[16] Z. Chen, D. Cummins, B. N. Reinecke, E. Clark, M. K. Sunkara, T. F. Jaramillo, *Nano Lett.* **2011**, *11*, 4168.
[17] G. Ye, Y. Gong, J. Lin, B. Li, Y. He, S. T. Pantelides, W. Zhou, R. Vajtai, P. M. Ajayan, *Nano Lett.* **2016**, *16*, 1097.
[18] D. Voiry, H. Yamaguchi, J. Li, R. Silva, D. C. B. Alves, T. Fujita, M. Chen, T. Asefa, V. B. Shenoy, G. Eda, M. Chhowalla, *Nat. Mater.* **2013**, *12*, 850.
[19] M. A. Lukowski, A. S. Daniel, F. Meng, A. Forticaux, L. Li, S. Jin, *J. Am. Chem. Soc.* **2013**, *135*, 10274.
[20] H. Li, C. Tsai, A. L. Koh, L. Cai, A. W. Contryman, A. H. Fragapane, J. Zhao, H. S. Han, H. C. Manoharan, F. Abild-Pedersen, J. K. Nørskov, X. Zheng, *Nat. Mater.* **2016**, *15*, 48.
[21] A.-Y. Lu, H. Zhu, J. Xiao, C.-P. Chuu, Y. Han, M.-H. Chiu, C.-C. Cheng, C.-W. Yang, K.-H. Wei, Y. Yang, Y. Wang, D. Sokaras, D. Nordlund, P. Yang, D. A. Muller, M.-Y. Chou, X. Zhang, L.-J. Li, *Nat. Nanotechnol.* **2017**, *12*, 744.
[22] M. Yagmurcukardes, Y. Qin, S. Ozen, M. Sayyad, F. M. Peeters, S. Tongay, H. Sahin, *Appl. Phys. Rev.* **2020**, *7*, 011311.
[23] D. Er, H. Ye, N. C. Frey, H. Kumar, J. Lou, V. B. Shenoy, *Nano Lett.* **2018**, *18*, 3943.
[24] J. Zhang, S. Jia, I. Kholmanov, L. Dong, D. Er, W. Chen, H. Guo, Z. Jin, V. B. Shenoy, L. Shi, J. Lou, *ACS Nano* **2017**, *11*, 8192.
[25] Z. Gan, I. Paradisanos, A. Estrada-Real, J. Picker, E. Najafidehaghani, F. Davies, C. Neumann, C. Robert, P. Wiecha, K. Watanabe, T. Taniguchi, X. Marie, J. Biskupek, M. Mundszinger, R. Leiter, U. Kaiser, A. V. Krasheninnikov, B. Urbaszek, A. George, A. Turchanin, *Adv. Mater.* **2022**, *34*, 2205226.
[26] M. M. Petrić, M. Kremser, M. Barbone, Y. Qin, Y. Sayyad, Y. Shen, S. Tongay, J. J. Finley, A. R. Botello-Méndez, K. Müller, *Phys. Rev. B.* **2021**, *103*, 035414.

[27] J. Shi, H. Xu, C. Heide, C. HuangFu, C. Xia, F. de Quesada, H. Shen, T. Zhang, L. Yu, A. Johnson, F. Liu, E. Shi, L. Jiao, T. Heinz, S. Ghimire, J. Li, J. Kong, Y. Guo, A. M. Lindenberg, *Nat. Commun.* **2023**, *14*, 4953.
[28] S. W. Kim, S. Y. Choi, S. H. Lim, E. B. Ko, S. Kim, Y. C. Park, S. Lee, H. H. Kim, *Adv. Funct. Mater.* **2024**, *34*, 2308709.
[29] J. Picker, M. Ghorbani-Asl, M. Schaal, S. Kretschmer, F. Otto, M. Gruenewald, C. Neumann, T. Fritz, A. V. Krasheninnikov, A. Turchanin, *Nano Lett.* **2025**, *25*, 3330.
[30] J. Schmeink, V. Musytschuk, E. Pollmann, S. Sleziona, A. Maas, P. Kratzer, M. Schleberger, *Nanoscale* **2023**, *15*, 10834.
[31] K. Rahimi, *Int. J. Hydrog. Energy* **2025**, *161*, 150744.
[32] S. Pakhira, S. N. Upadhyay, *Sustainable Energy & Fuels* **2022**, *6*, 1733.

# Supporting Information

## Sonochemically Boosted Hydrogen Evolution Activity of Janus TMD Monolayers

*Rayantan Sadhukhan[1†], Md Tarik Hossain[2†], Julian Picker[2], Mahdi Ghorbani-Asl[3], Christof Neumann[2], Arkady V. Krasheninnikov[3], Tharangattu N. Narayanan[1]*, Andrey Turchanin[2,4]**

[1]Tata Institute of Fundamental Research Hyderabad,
Serilingampally Mandal, Hyderabad, Telangana 500 046, India.
[2]Friedrich Schiller University Jena, Institute of Physical Chemistry,
Lessingstr. 10, 07743 Jena, Germany.
[3]Helmholtz-Zentrum Dresden-Rossendorf, Institute of Ion Beam Physics and Materials Research,
Bautzner Landstr. 400, 01328 Dresden, Germany.
[4]Center for Energy and Environmental Chemistry, Philosophenweg 7a, 07743 Jena, Germany.

*[†] = Contributed equally*

Present address for J.P.: MAX IV Laboratory, Lund University, Fotongatan 2, 224 84 Lund

**Corresponding Authors**:

Tharangattu N. Narayanan (tnn@tifrh.res.in)

Andrey Turchanin (andrey.turchanin@uni-jena.de)

**Materials and Methods**

*Growth of $MoS_2$, $MoSe_2$, $WS_2$ and $WSe_2$ on Au foil:*

Monolayer $MoS_2$, $MoSe_2$, $WS_2$, and $WSe_2$ were synthesized on Au foil (99.95 %, thickness ~0.025 mm, size ~ 1 × 1 $cm^2$, Alfa Aesar) using a modified chemical vapor deposition (CVD) process adapted from a previously reported method.[1] Growth was carried out in a two-zone split-tube furnace (Carbolite Gero; 55 mm inner diameter) equipped with an inner quartz tube (15 mm inner diameter) to confine precursor transport. Prior to the growth, Au foils were cleaned by $O_2$ plasma, immersed in isopropanol for 30 s, dried under $N_2$, and annealed at 1040 °C in flowing Ar.

In all cases, the chalcogen precursor (elemental sulfur or selenium, Sigma Aldrich) was loaded into a quartz Knudsen cell and placed at the center of zone 1, while the transition metal oxide precursor ($MoO_3$ or $WO_3$) with a 0.1 mg $mL^{-1}$ NaCl aqueous droplet was deposited onto a $SiO_2$/Si substrate and placed at the center of zone 2. The growth chamber was evacuated and backfilled with Ar to atmospheric pressure, followed by growth under 100 $cm^3$ $min^{-1}$ Ar flow. Zone 2 was heated to the specified growth temperature at a rate of 40 °C $min^{-1}$, while zone 1 was maintained at the specified chalcogen temperature. Growth conditions were held for 15 minutes to achieve maximum monolayer coverage. Growth parameters are tabulated below:

| Material | Chalcogen source | Transition metal source | Zone 2 (°C) | Zone 1 (°C) |
|---|---|---|---|---|
| $MoS_2$ | Sulfur powder (99.98%) | $MoO_3$ (≥99.5%) | 770 | 200 |
| $MoSe_2$ | Selenium granules (≥99.999%) | $MoO_3$ (≥99.5%) | 770 | 200 |
| $WS_2$ | Sulfur powder (99.98%) | $WO_3$ (99.998%) | 820 | 420 |
| $WSe_2$ | Selenium granules (≥99.999%) | $WO_3$ (99.998%) | 820 | 420 |

*Growth of SeMoS and SeWS:*

For the synthesis of Janus SeMoS and SeWS, monolayer $MoSe_2$ and $WSe_2$, respectively, underwent post-growth chalcogen exchange by exposure to sulfur vapor at 700 °C, following a previously reported procedure.[2] The process was performed in the same CVD system, with both furnace zones heated simultaneously at a ramp rate of 40 °C $min^{-1}$ to 200 °C (zone 1, sulfur source) and 700 °C (zone 2, sample), and held at these temperatures for 15 min. The temperature profile for growing these monolayers is shown in the supplementary Fig. S1.

*Solvent treatment protocol:*

In the first protocol, monolayers were immersed in various solvents - polar (water, isopropyl alcohol (IPA), and chloroform) and non-polar (benzene) - for 12 hours. The second protocol involved heating the immersed samples at 90 °C for 5 hours. In the third protocol, a sonochemical treatment, the solvent-immersed monolayers were sonicated for 30 minutes using a laboratory bath sonicator operating at 40 kHz (Branson). All treated Janus and conventional TMD monolayers were then tested for their electrocatalytic HER performance in acidic medium (0.5 M $H_2SO_4$, pH ~ 0).

*Characterization:*

Optical microscopy: Bright-field optical images were acquired on a Zeiss Axio Imager Z1.m microscope equipped with a thermoelectrically cooled 3-megapixel CCD camera (Axiocam 503 color).

Raman spectroscopy: Raman spectra were obtained using a Renishaw in-Via microscope with 532 nm (Nd: YAG) excitation through a 50× long-working-distance objective and a 2400 lines $mm^{-1}$ grating, with a spacing of ~ 0.87 $cm^{-1}$ between consecutive spectral data points. Signals were collected with a CCD detector. The Si Raman peak at 521 $cm^{-1}$ was used for calibration. Data was analyzed using Wire 5.6 software.

X-ray photoelectron spectroscopy (XPS): XPS was conducted using a monochromatic Al $K_\alpha$ X-ray source (1486.7 eV) and an electron analyzer with a spectral energy resolution of 0.6 eV (Argus CU). For angle-resolved XPS (ARXPS) measurements, the photoelectron emission angle (θ) was varied between 0° and 80°, while the source–analyzer angle was maintained at 54.7°. For analyzing the XP spectra, the software CasaXPS was employed. All spectra were calibrated to the Au $4f_{7/2}$ peak at a binding energy (BE) of 84.0 eV. Peaks were fitted by Voigt functions (Gaussian: Lorentzian = 70:30) after subtraction of a linear (S 2p, Se 3d) or Shirley (Mo 3d, S 2s, Se 3s, W 4f) background. For quantification of elemental ratios, Scofield's relative sensitivity factors of 1.11, 1.36, 5.48, 1.43, and 5.62 were used for the S $2p_{3/2}$, Se $3d_{5/2}$, W $4f_{7/2}$, S 2s/Se 3s, and Mo $3d_{5/2}$ orbitals, respectively.

Scanning electron microscopy (SEM): SEM images were acquired on a Zeiss Sigma VP field-emission microscope operated at 10 kV with an in-lens detector.

Thin film X-ray diffraction (XRD): XRD experiments were performed in RIGAKU Smartlab X-Ray diffractometer using Cu Kα radiation (λ = 1.4518 Å) at a scan rate of 1 °/minute.

*Electrochemical measurement:*

All the electrochemical measurements were performed in a Biologic SP 300 instrument. The experiments were done in 0.5M $H_2SO_4$ (pH = 0) using Janus on Au foil as the working electrode, Ag/AgCl ($KCl_{sat}$) as reference electrode, and Graphite rod as the counter electrode (as shown in supplementary Fig. S4). Later, all the potentials were converted to the reversible hydrogen electrode (RHE) scale according to Equation 1.

$$E_{RHE} = E_{Ag/AgCl} + 0.197 + 0.0591 \text{ x pH (where pH} \sim 0) \quad (1)$$

All the experiments were performed at room temperature, and the electrode activation was done by cycling the electrode between 0 and - 0.5 V vs Ag/AgCl ($KCl_{sat}$) for 15 cycles at 200 $mVs^-$

$^{1}$. The Linear Sweep Voltammetry (LSV) measurements are performed at a scan rate of 10 $mVs^{-1}$, and the Cyclic Voltammetry (CV) measurements are performed at 200 $mVs^{-1}$.

*Electrochemical surface area calculation:*

Electrochemical surface area (ECSA) for Au were determined from CV by measuring the area of the oxidation region of a gold oxide monolayer in 0.5M $H_2SO_4$ (acidic) electrolyte. The area was then divided by the scan rate to obtain the total charge ($Q_{tc}$) involved in monolayer oxidation of Au. The specific charge ($Q_{sc}$) for Au is 384 μC/cm². Hence, ECSA is calculated following Equation 2.

$$ECSA = Q_{tc}/Q_{sc} \quad (2)$$

*EIS measurement details:*

Electrochemical Impedance Spectroscopy measurements are also performed in a SP 300 potentiostat, scanning from 1 MHz to 10 mHz frequency range. The resultant EIS plot is then fitted in a Randle's circuit using EC-Lab software V11.50. The equivalent Randle's circuit is shown in the supplementary Fig. S8.

*Tafel slope and exchange current density calculation:*

The Tafel slopes were calculated using the LSV plots, and the slope was measured around the onset region of all the catalysts (± 50 mV). For the calculation of overpotential, we considered -0.197 V vs Ag/AgCl ($KCl_{sat}$) as the 0 V overpotential point for the Tafel slope calculation. Exchange Current densities are the current for a specific reaction without providing any overpotential. So the Tafel slope plot is extended till y (mV) = 0 (which is 0 overpotential), to know the exchange current density, which is plotted on the x-axis, as shown in the supplementary Fig. S9.

*Computational Details:*

First-principles calculations were carried out within the framework of the density functional theory (DFT) using the projector augmented-wave (PAW) method, as implemented in the VASP package.[3, 4] Exchange–correlation effects were described by the Perdew–Burke–Ernzerhof (PBE) functional within the generalized gradient approximation[5]. A kinetic-energy cutoff of 450 eV was used for the plane-wave expansion.

To eliminate spurious interactions between periodic images, a vacuum spacing of nearly 20 Å was introduced along the out-of-plane direction. Structural relaxations were continued until the residual force on each atom became smaller than 0.01 eV/Å. Owing to the large sizes of the supercells, only Γ-point calculations were carried out. Dispersion corrections were incorporated through Grimme's DFT-D3 scheme to properly account for long-range van der Waals interactions,[6]

For the SeMoS/Au(111) heterostructure, the interface was modelled using a 10×10 Au(111) slab combined with a 9×9 SeMoS monolayer supercell, resulting in a minimal lattice mismatch of approximately 0.13%. The heterostructure contained a total of 643 atoms.

The adsorption energy ($E_{\mathrm{ads}}$) was determined from

$$E_{ads} = E_{\mathrm{slab+adsorbate}} - E_{\mathrm{slab}} - E_{\mathrm{adsorbate}}$$

where the $E_{\mathrm{slab+adsorbate}}$ denotes the total energy of the adsorbate–substrate system, while $E_{\mathrm{slab}}$ and $E_{\mathrm{adsorbate}}$ correspond to the isolated substrate and adsorbate, respectively.

*Gibbs Free-Energy Analysis:*

The HER mechanism on Janus MoSSe and Janus WSSe has been shown to follow the Volmer–Heyrovsky pathway,[7, 8] The Gibbs free energy of hydrogen adsorption (ΔG) was calculated to evaluate the catalytic performance of the investigated systems toward the hydrogen evolution reaction (HER). The free-energy variation was determined using the computational hydrogen electrode (CHE) method as follows:[9]

$$\Delta G_H = \Delta E_H + \Delta E_{ZPE} - T\Delta S$$

where $\Delta E_H$ is the hydrogen adsorption energy obtained from DFT calculations, $\Delta E_{ZPE}$ is the zero-point energy correction, and $T\Delta S$ corresponds to the entropy contribution at room temperature. Following the standard hydrogen electrode model, the entropy correction for adsorbed hydrogen was approximated using half of the entropy of $H_2$ gas.

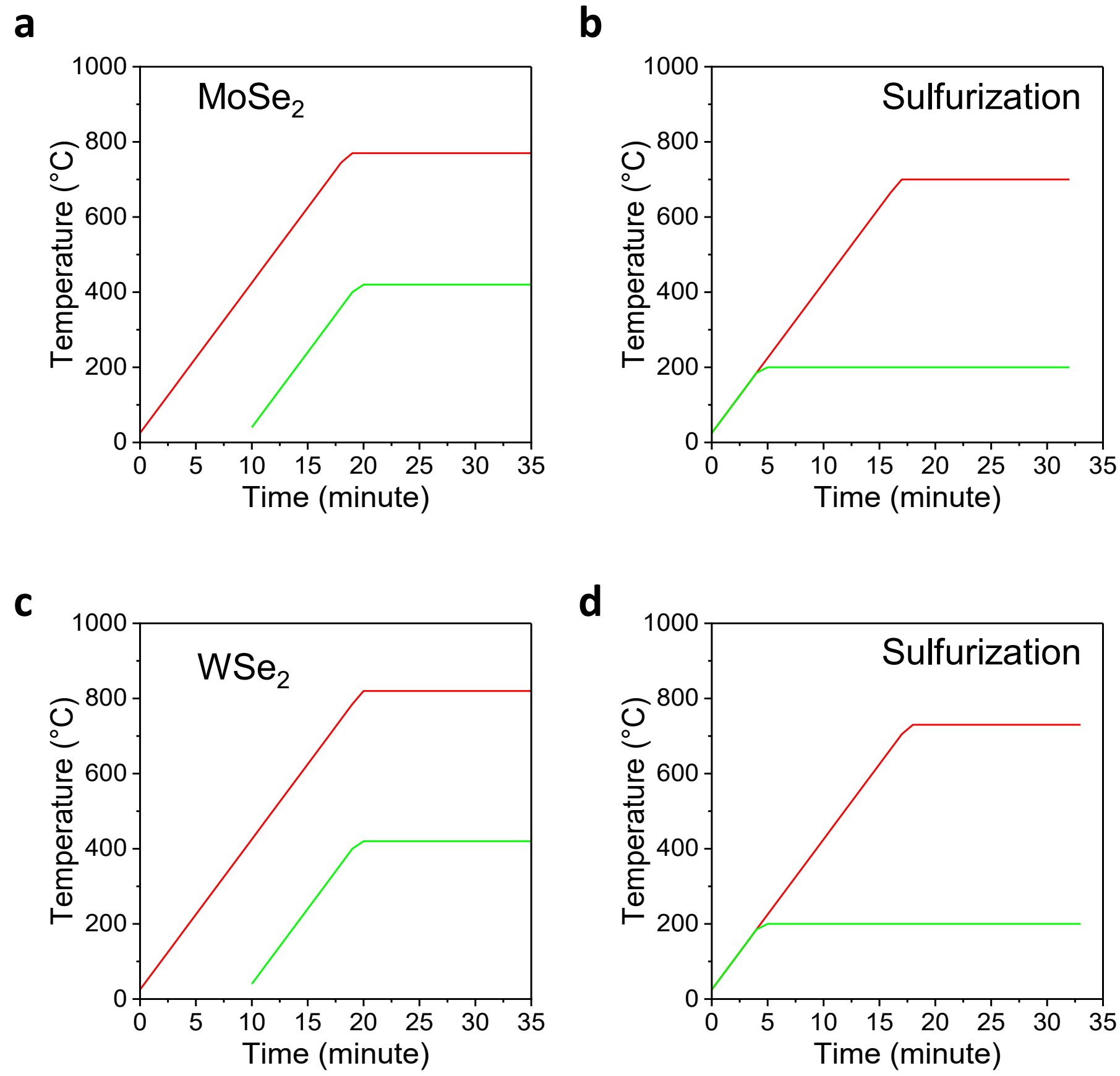

**Figure S1: Temperature profile for growing monolayer SeMoS and SeWS.** (a) $MoSe_2$ growth, (b) sulfurization of $MoSe_2$ for Janus SeMoS conversion, (c) $WSe_2$ growth, and (d) sulfurization of $WSe_2$ for Janus SeWS conversion.

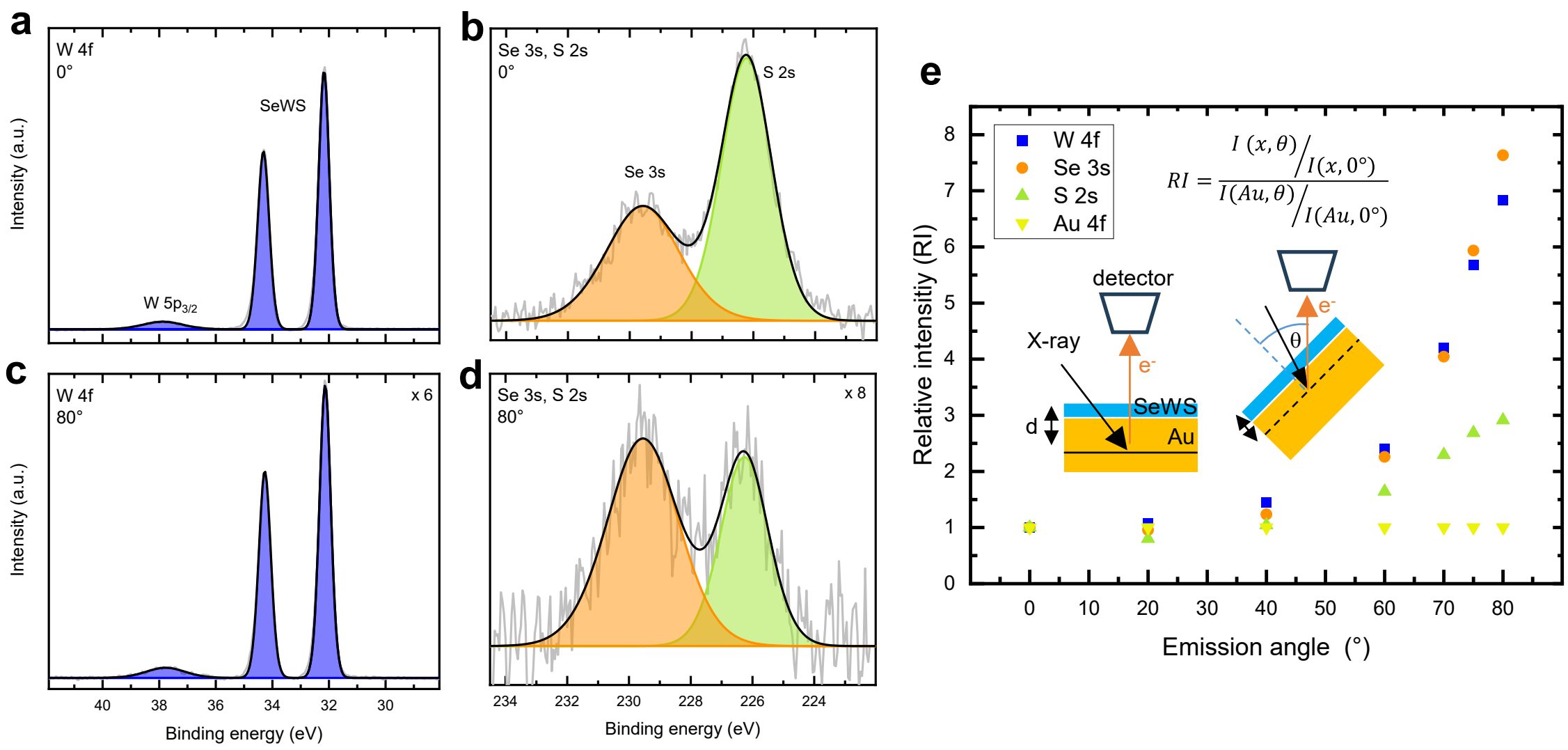

$$RI = \frac{I(x,\theta)/I(x,0°)}{I(Au,\theta)/I(Au,0°)}$$

**Figure S2: XPS and ARXPS characterization of SeWS.** (a-d) High-resolution W 4f, and S 2s/Se 3s XP spectra of as-grown Janus SeWS on Au foil measured at an emission angle (θ) of 0° (normal emission, top) and 80° (bottom), respectively. e) Relative intensities (RIs) of the Janus SeWS elements (W 4f, S 2s, and Se 3s peaks) as well as the substrate reference Au 4f peak as a function of θ.

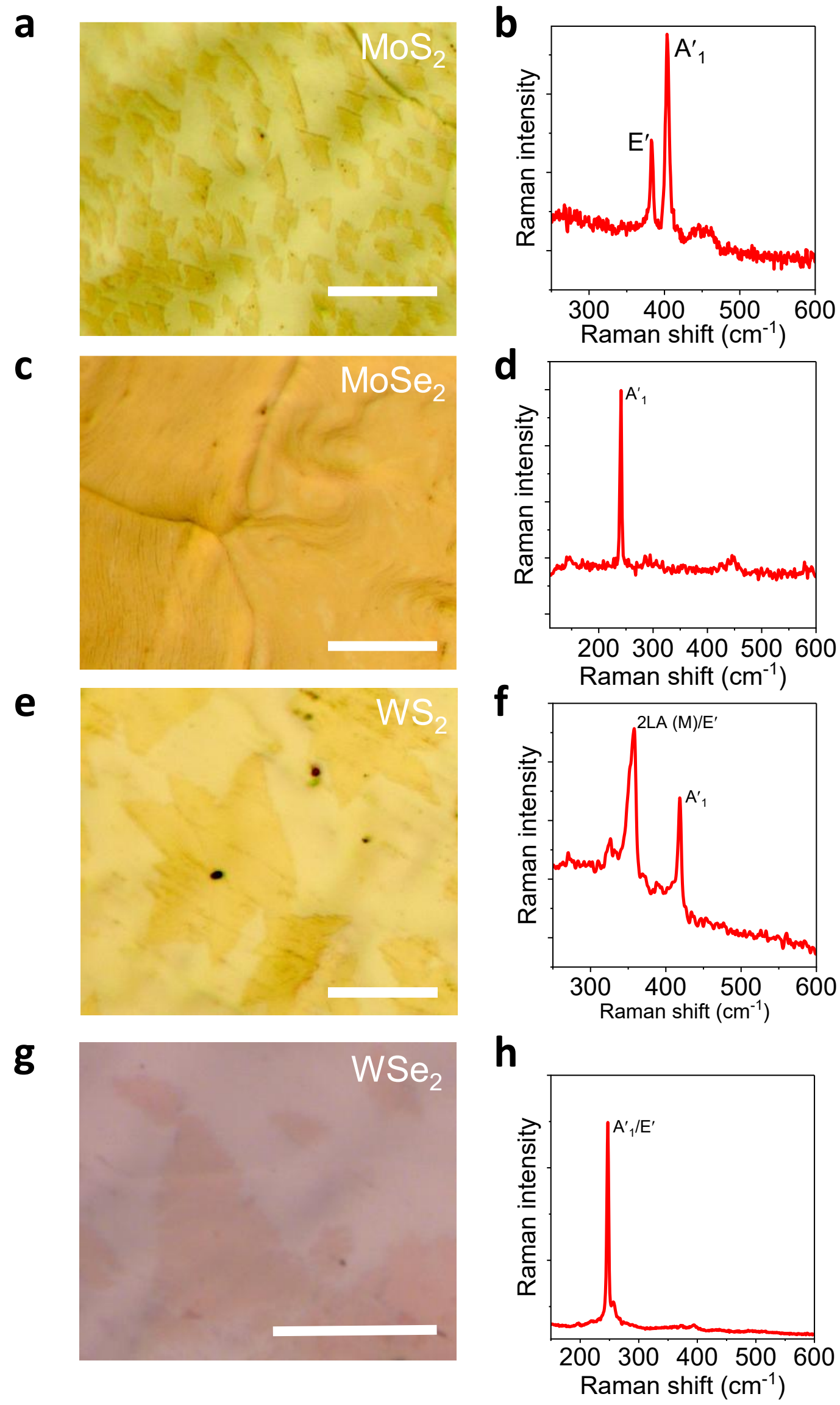

**Figure S3:** False color optical microscopy images of (a) $MoS_2$, (c) $MoSe_2$, (e) $WS_2$, and (g) $WSe_2$ on Au foil. Scale bar is 50 μm. (b, d, f, h) show the corresponding Raman spectrum of each TMD.

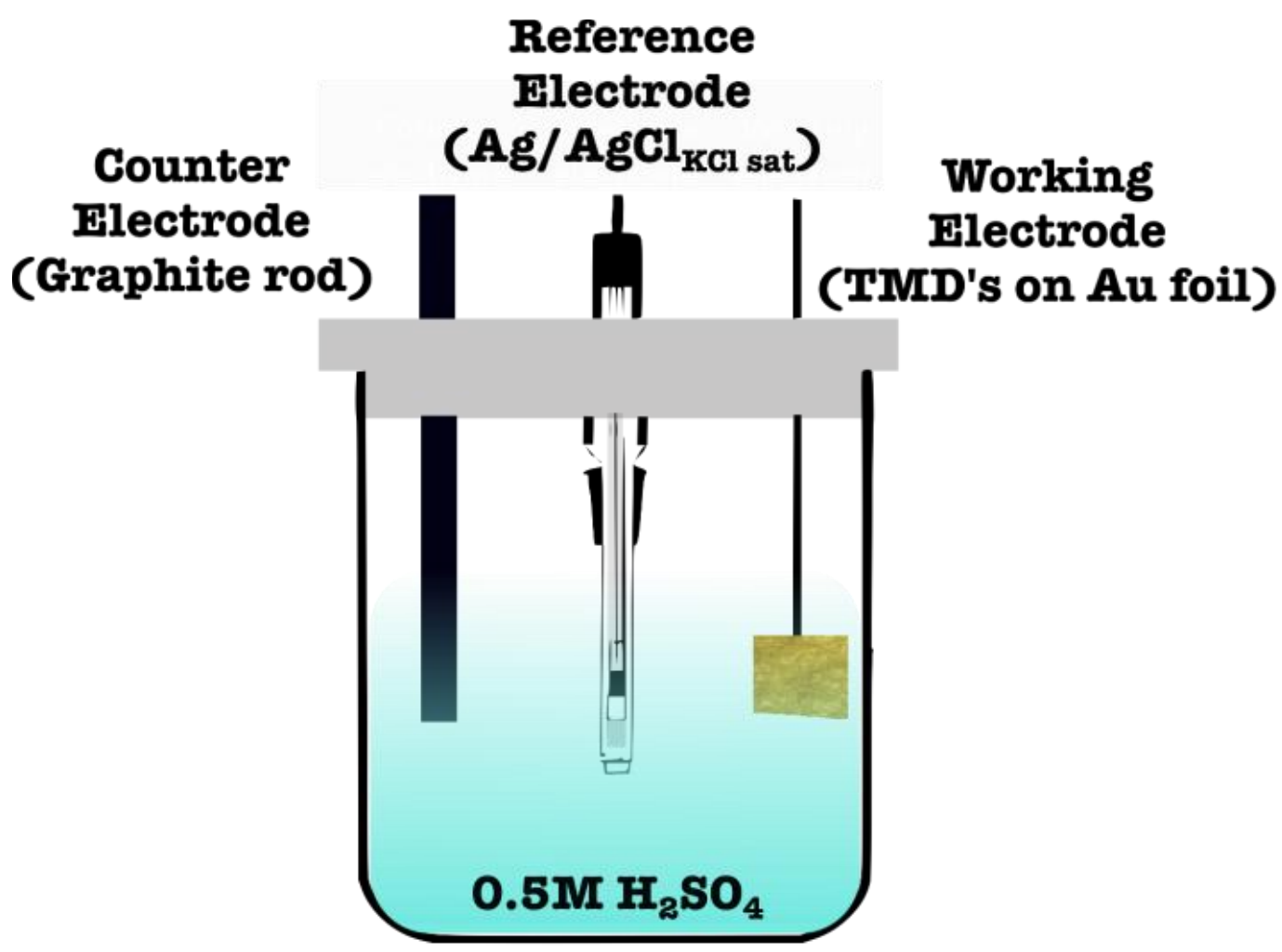

**Figure S4:** Schematic diagram of the electrochemical cell, where TMDs on Au foil work as the working electrode, Ag/AgCl ($KCl_{sat}$) works as a reference electrode, and the graphite rod works as a counter electrode.

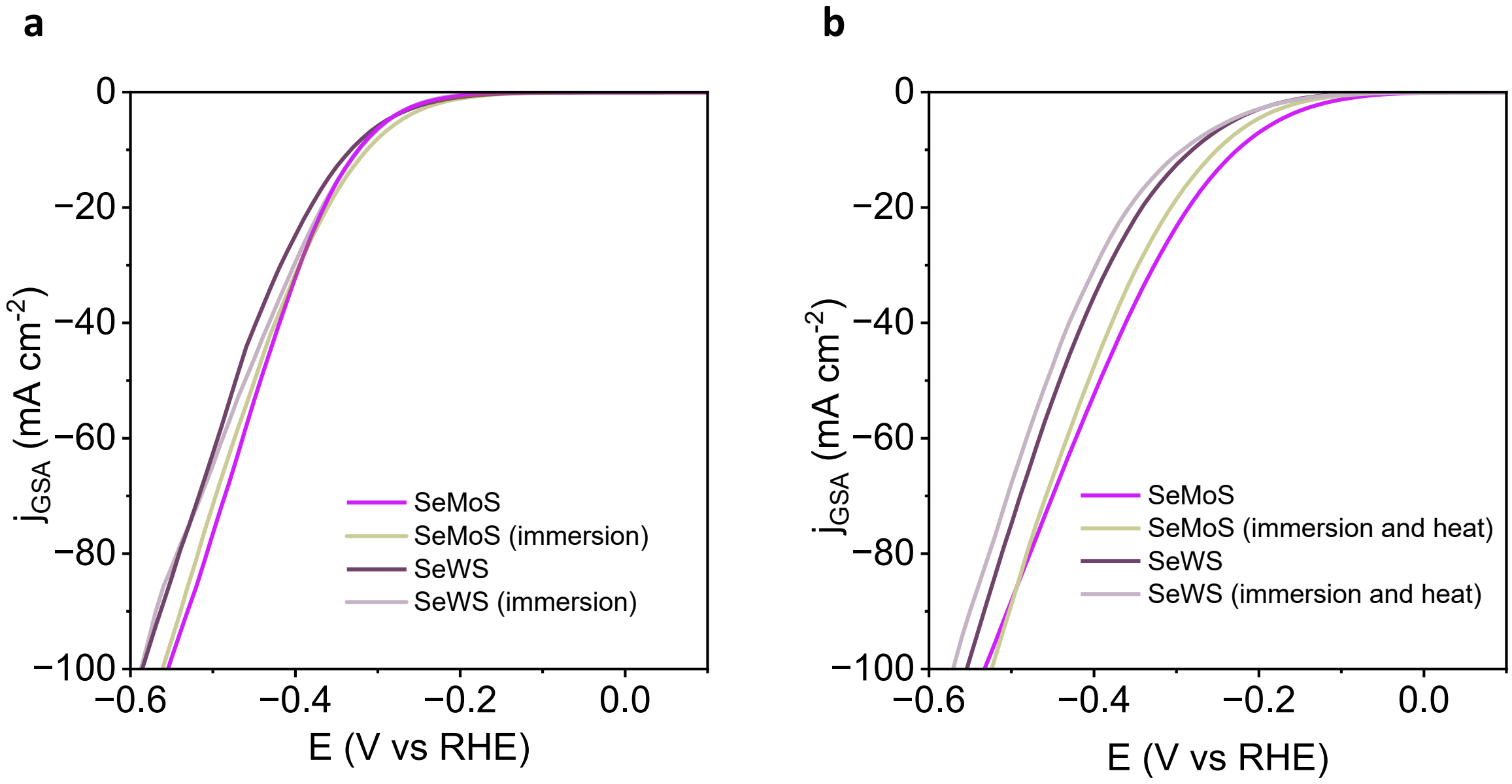

**Figure S5:** (a) Catalytic activity of the Janus samples after overnight immersion in water, showing no measurable degradation. (b) Catalytic activity after immersion in water and heating to 90 °C, also exhibiting no detectable change, demonstrates the stability of the samples under these conditions.

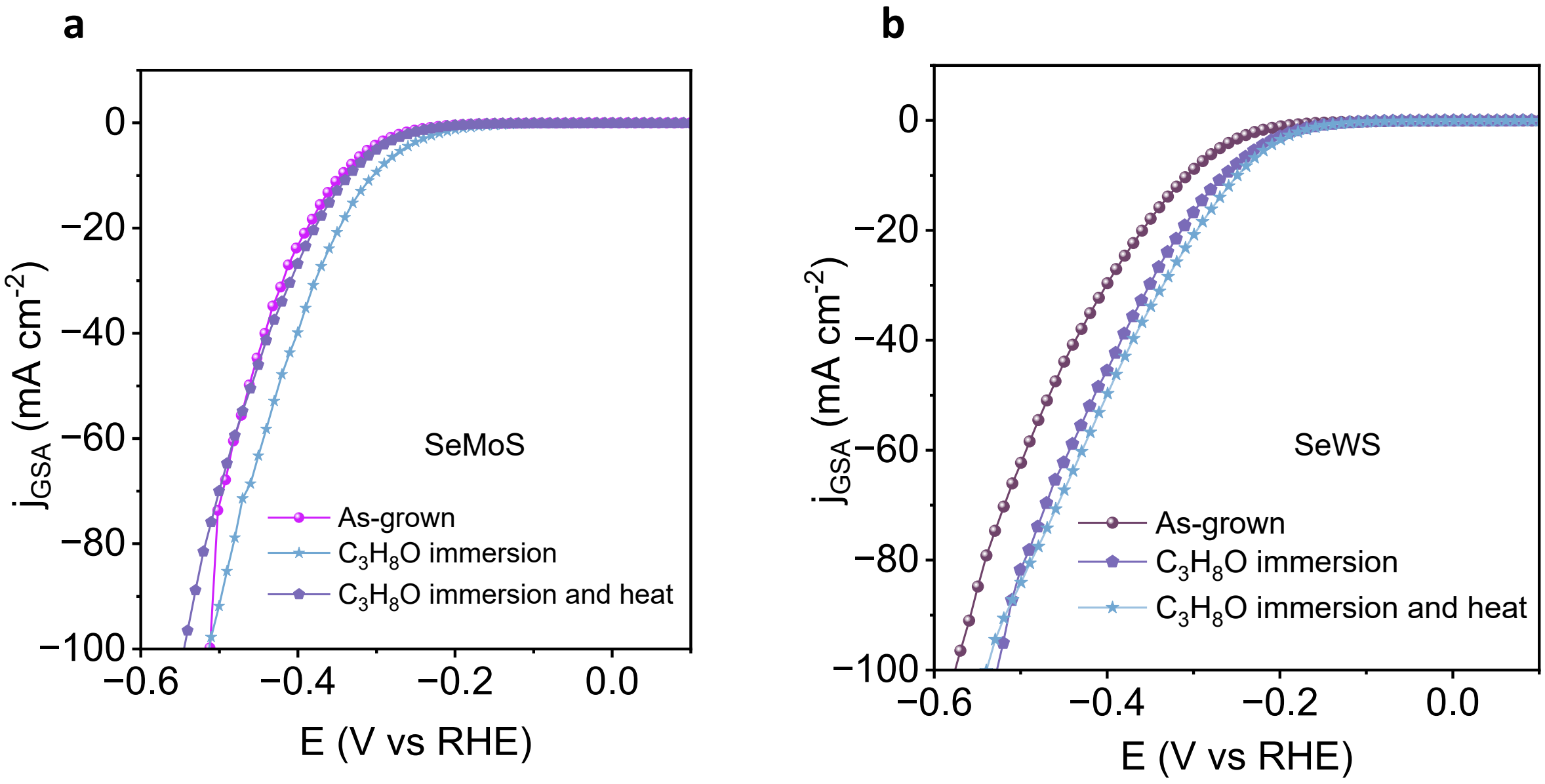

**Figure S6:** (a) Catalytic activity of the Janus samples after overnight immersion in IPA, showing no measurable enhancement or degradation. (b) Catalytic activity after immersion in IPA and heating to 70 °C, also exhibiting no detectable change, demonstrates the stability of the samples under these conditions.

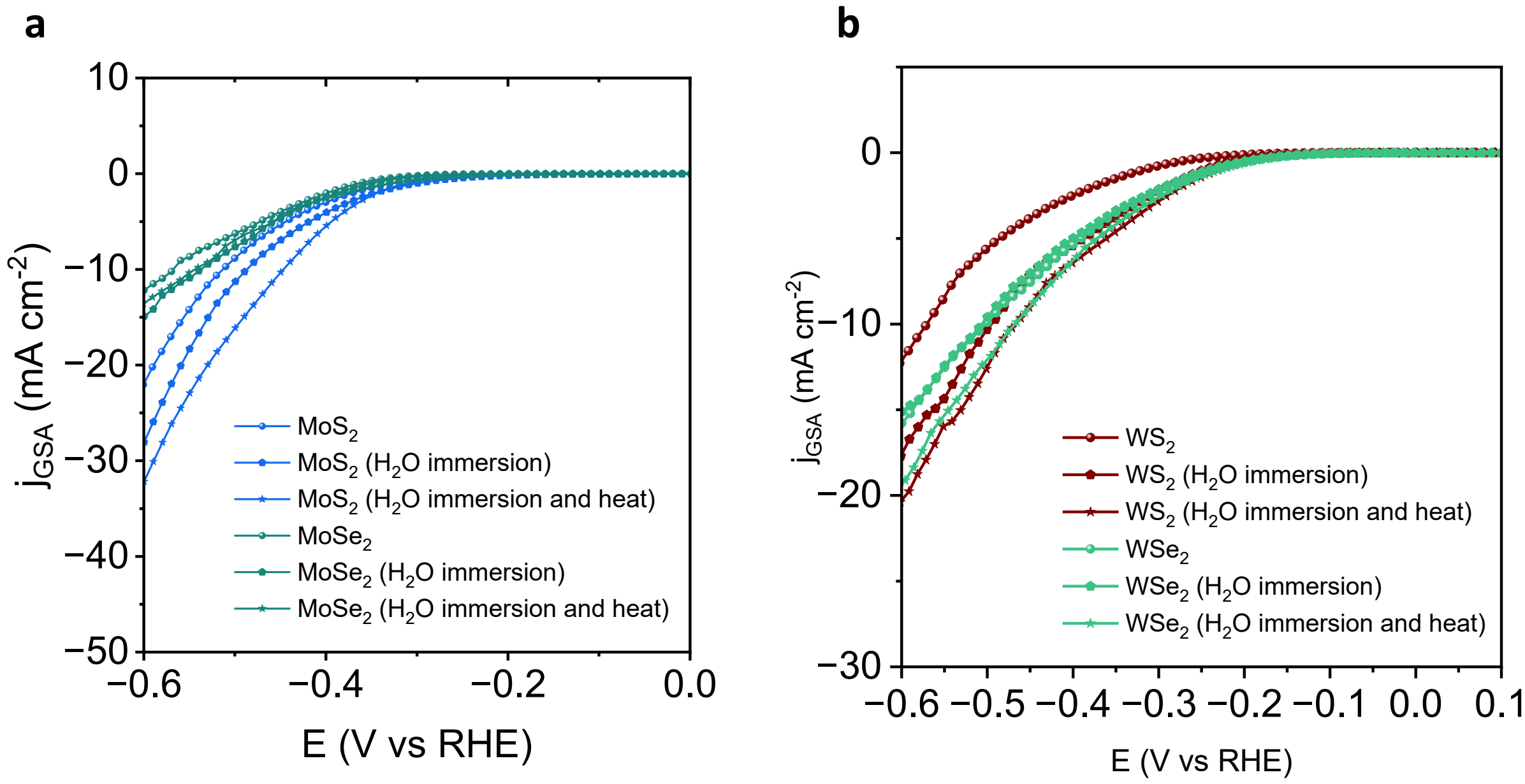

**Figure S7:** (a) Catalytic activity of the $MoS_2$ and $MoSe_2$ samples after overnight immersion in water or even when immersed in water and heated to 90 °C, showing no measurable enhancement. (b) Similar catalytic activity was checked for $WS_2$ and $WSe_2$, also exhibiting no detectable change.

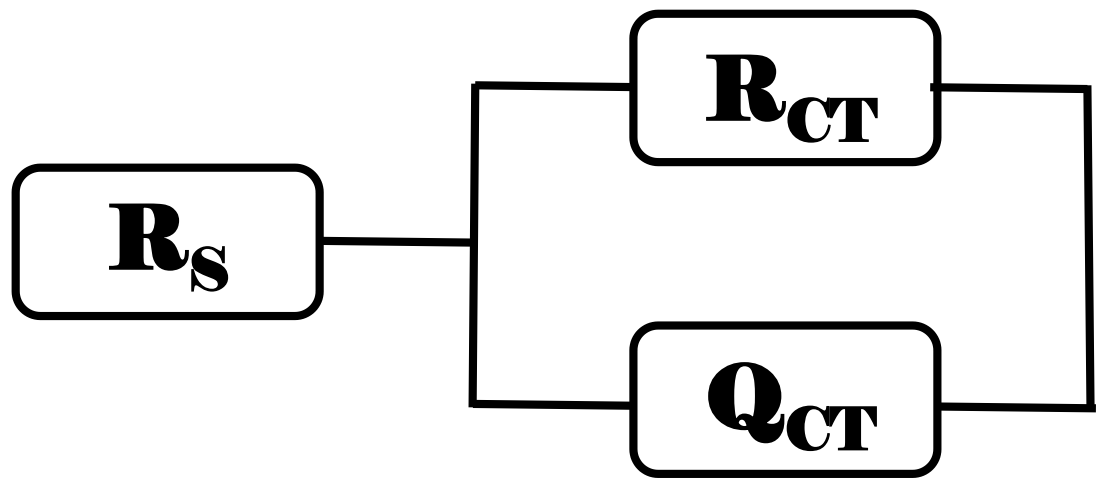

**Figure S8:** Equivalent Randles circuit for the electrochemical impedance spectroscopy, where $R_S$ is the solution resistance, $R_{CT}$ is the charge transfer resistance, and $Q_{CT}$ is the double layer capacitance with a phase element.

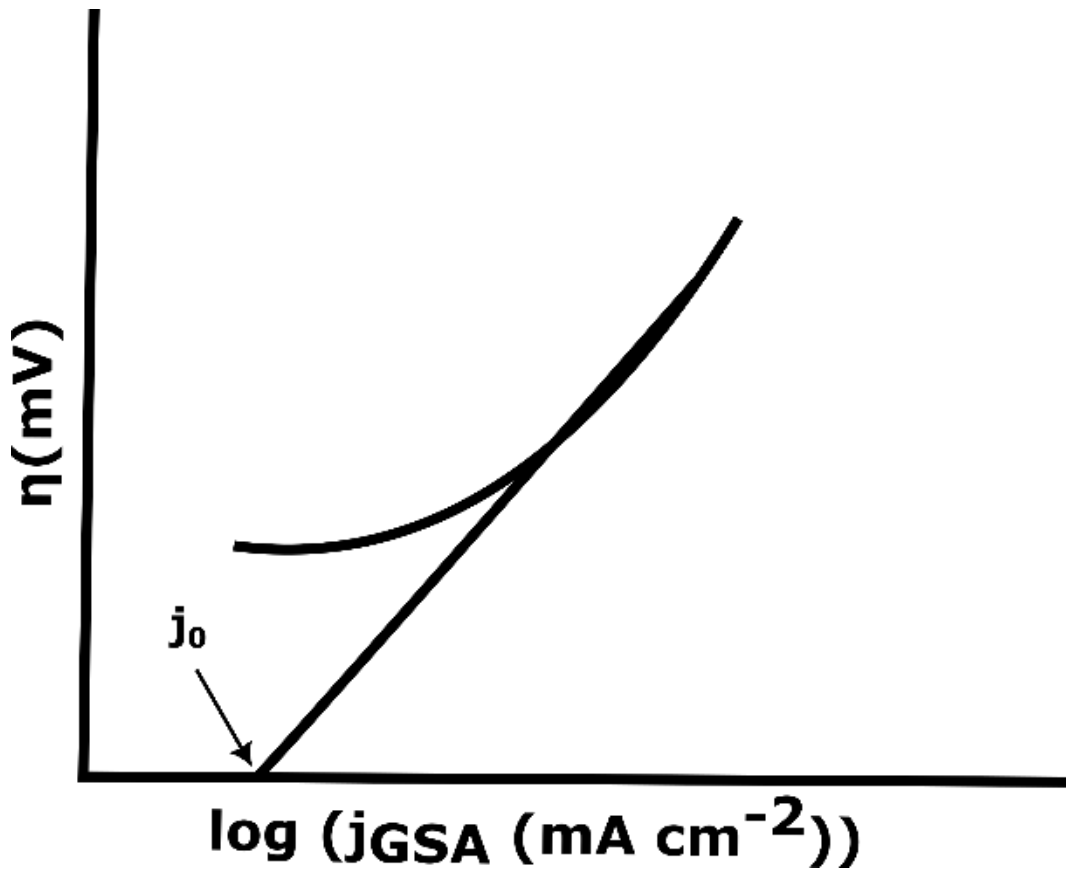

**Figure S9:** The Tafel slope plot is extended till y (mV) = 0 to know the exchange current density.

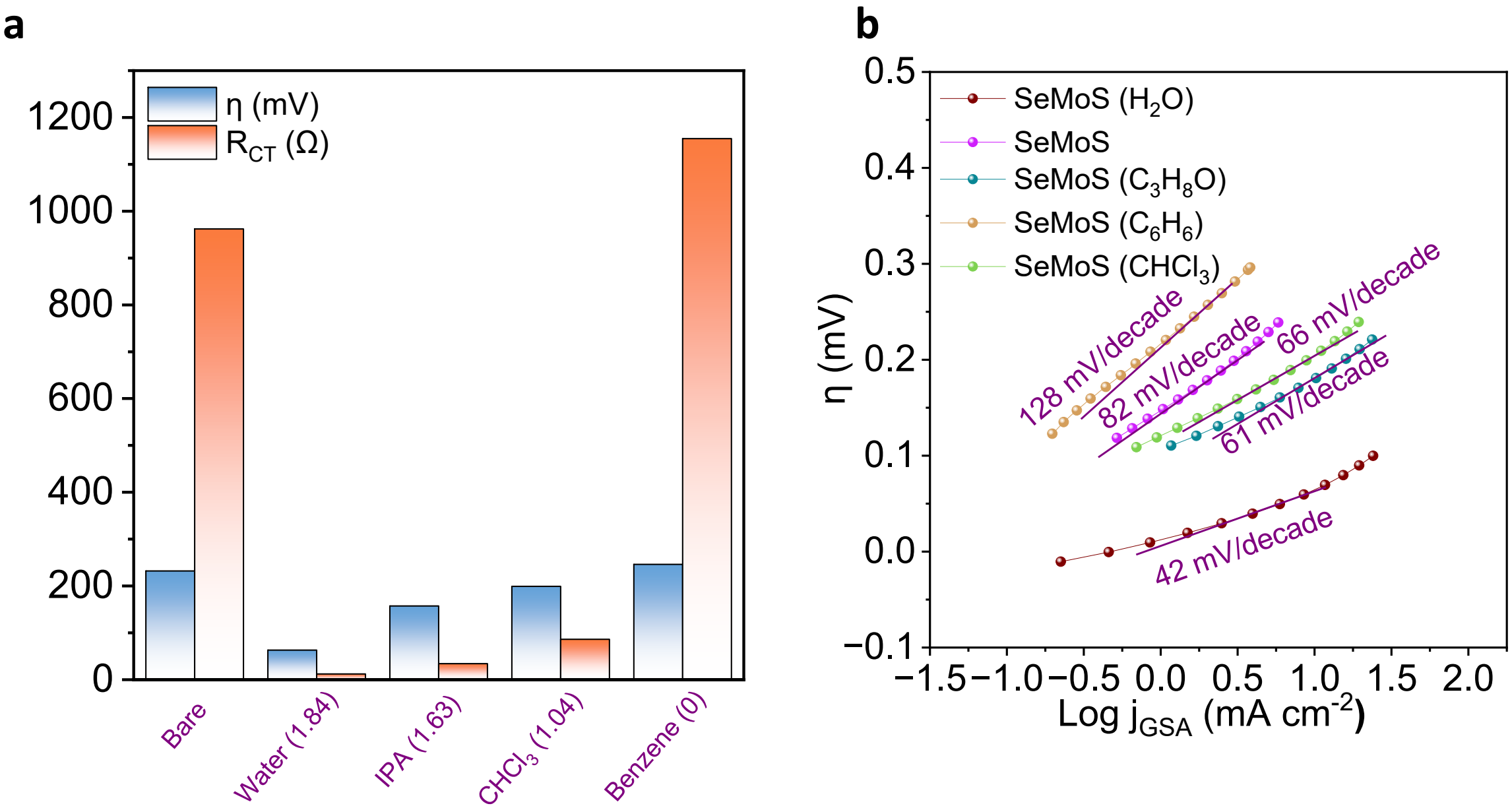

**Figure S10:** (a) Comparison of SeMoS Janus catalytic activity overpotentials ($\eta_{10}$) for HER and charge transfer resistances ($R_{CT}$) sonicated in different solvents, (b) Tafel slope plot for all the treated and untreated SeMoS Janus electrocatalyst.

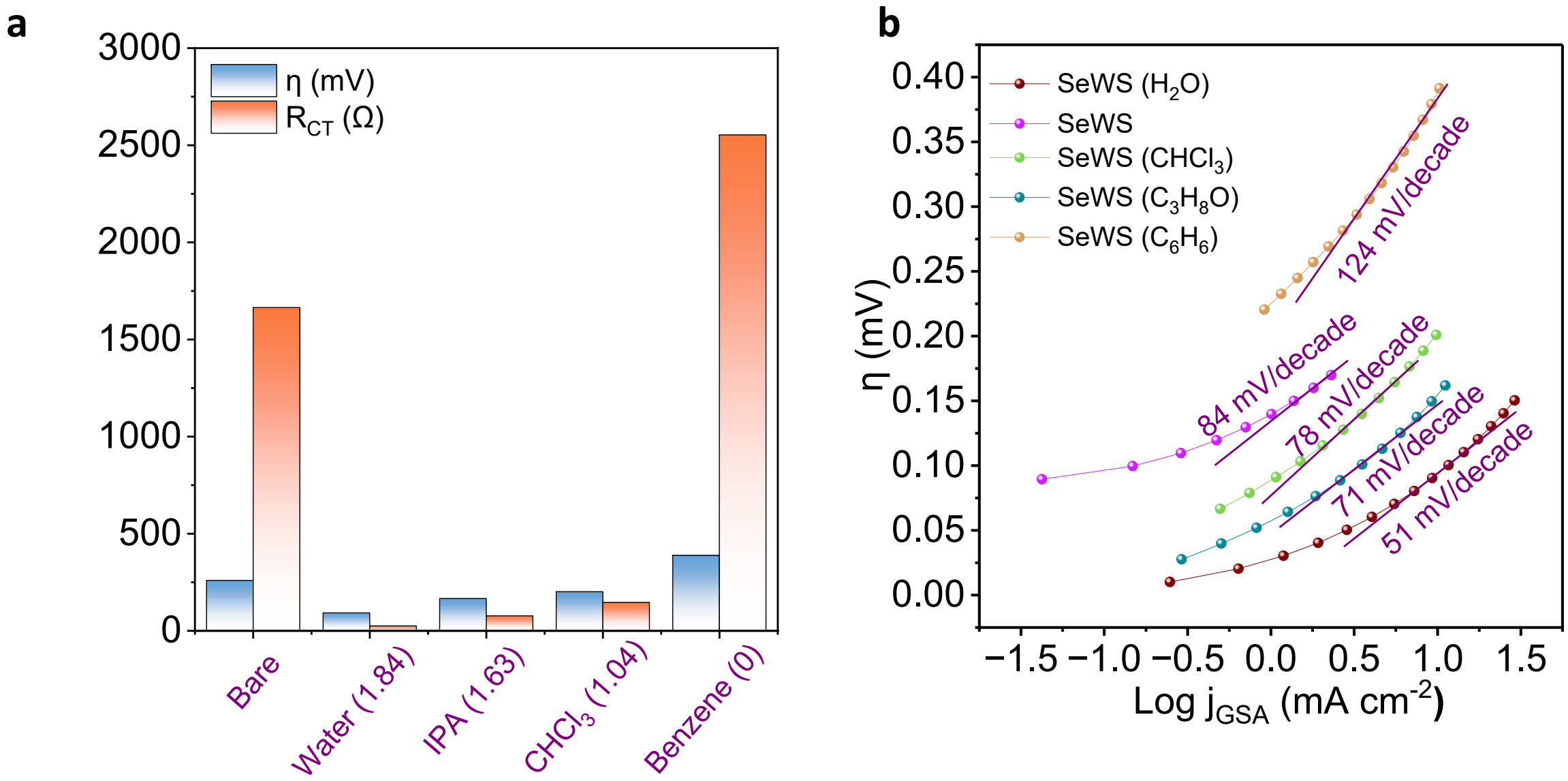

**Figure S11:** (a) Comparison of SeWS Janus catalytic activity overpotentials ($\eta_{10}$) for HER and charge transfer resistances ($R_{CT}$) sonicated in different solvents. (b) Tafel slope plot for all the treated and untreated SeWS Janus electrocatalyst.

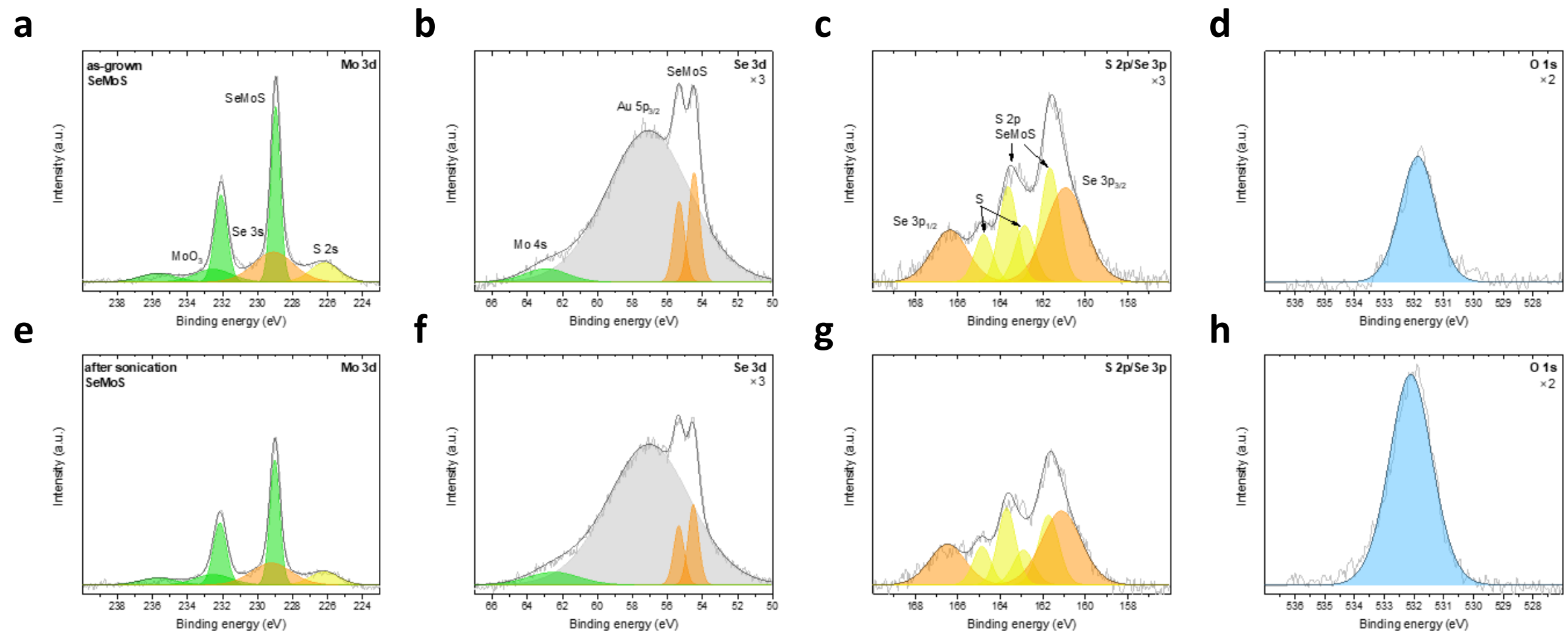

**Figure S12: XPS characterization of Janus SeMoS before and after sonochemical treatment.** High-resolution XP (a) Mo 3d, (b) Se 3d, (c) S 2p / Se 3p, and (d) O 1s spectra of as-grown SeMoS on Au foil. Panels (e-h) show the corresponding XP spectra after sonication in $H_2O$.

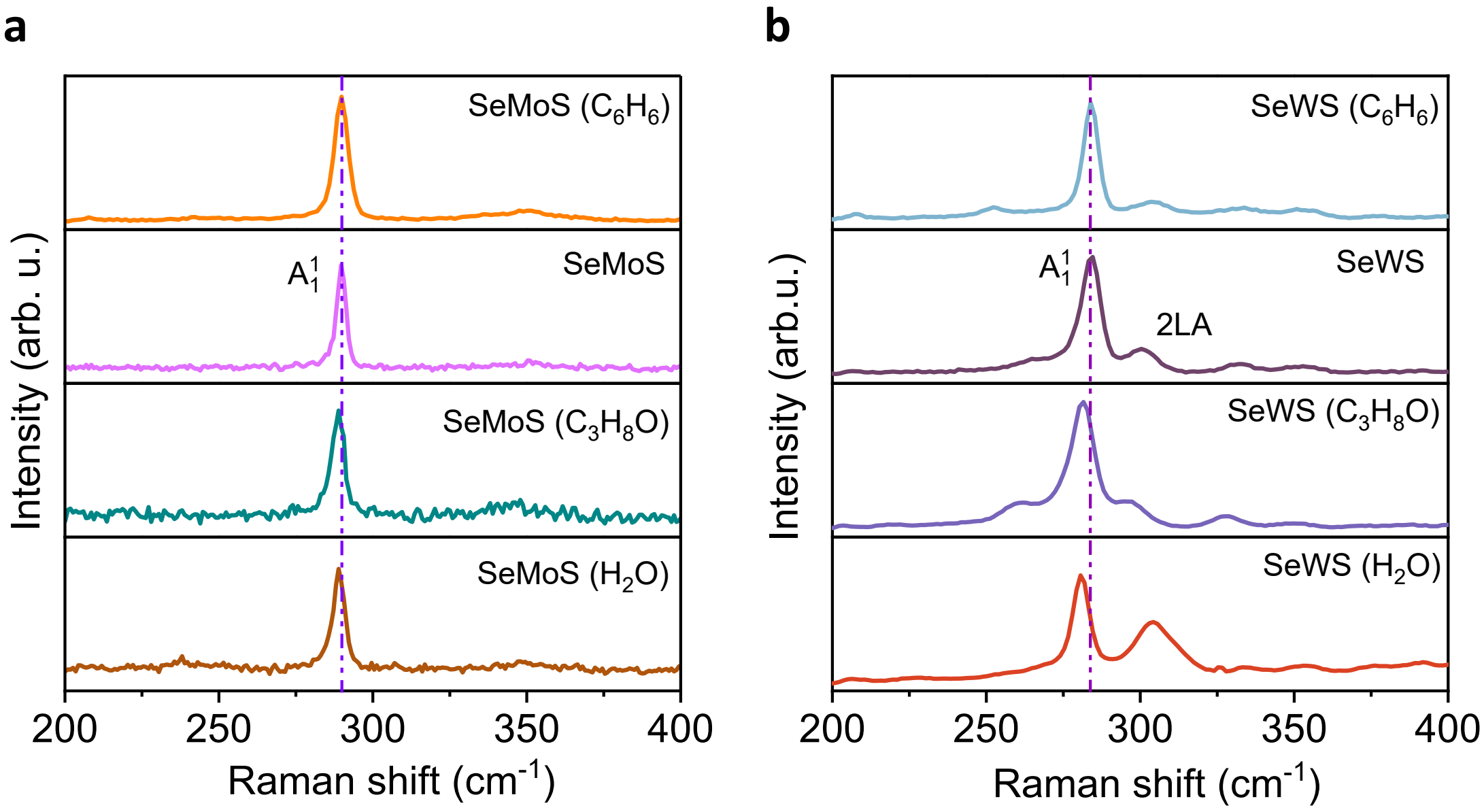

**Figure S13:** Raman spectra of the as-grown Janus TMDs and after treatment in different solvents showing Janus peaks in all the cases.

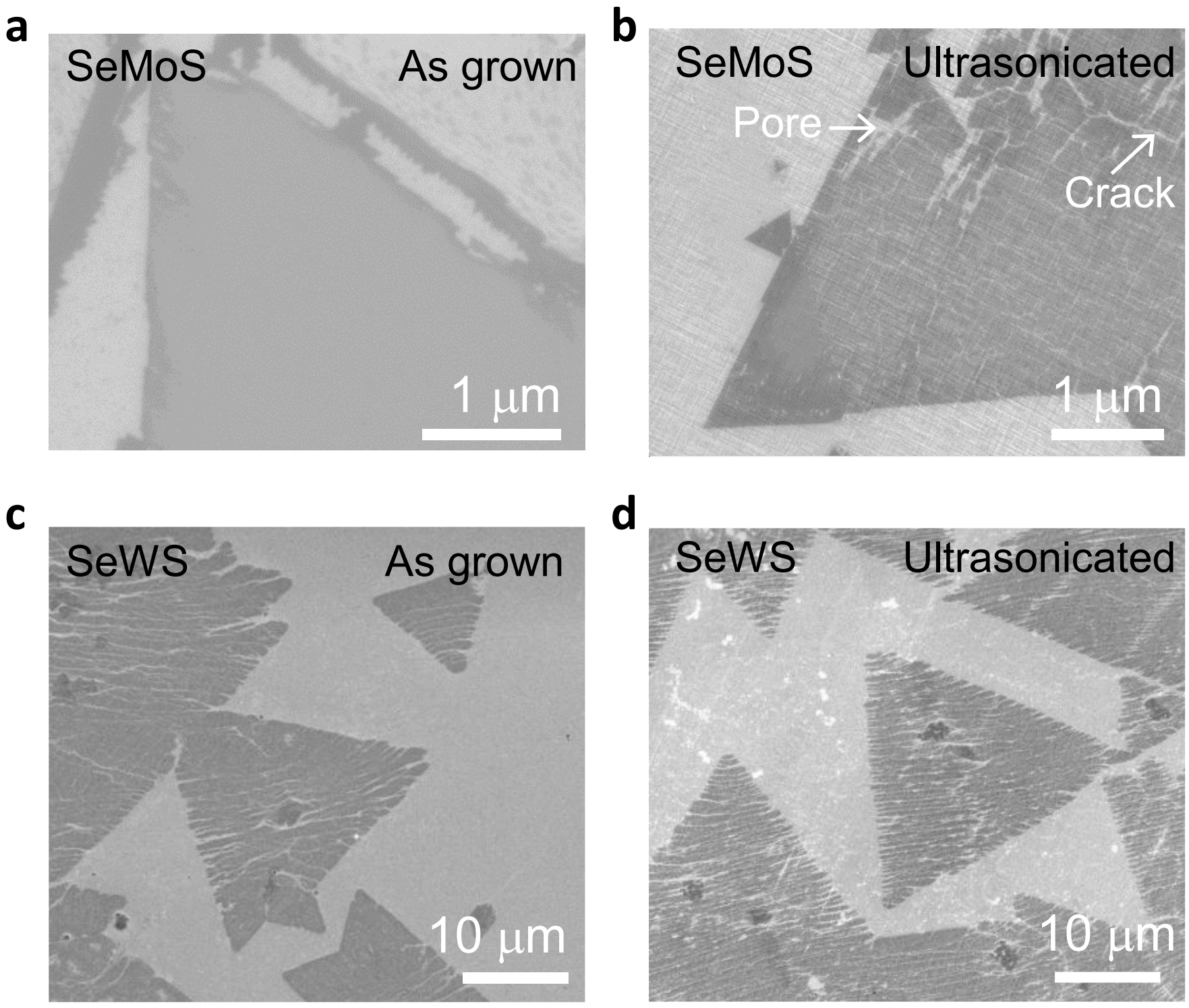

**Figure S14: Morphological changes in SeMoS and SeWS before and after water sonication.** SEM images of SeMoS (a,b) and SeWS (c,d) before (a,c) and after (b,d) water sonication. High-density nanometer-scale cracks and pores formed within the Janus layers after polar solvent treatment.

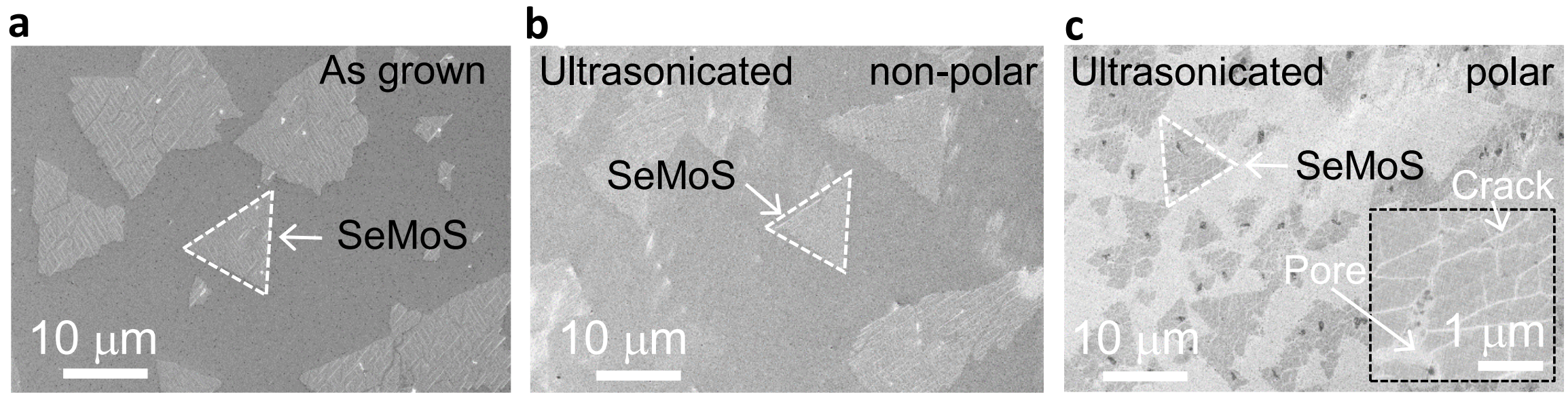

**Figure S15: Morphological changes in SeMoS under polar and nonpolar sonochemical treatment.** SEM images of SeMoS (a) as-grown, (b) after sonication in a polar solvent, and (c) after sonication in a non-polar solvent, highlighting the solvent-dependent changes in surface morphology. The inset in (c) shows a magnified view of cracks in Janus ML. High-density nanometer-scale cracks and pores formed within the SeMoS layers, with abundant exposed edges under polar solvent treatment.

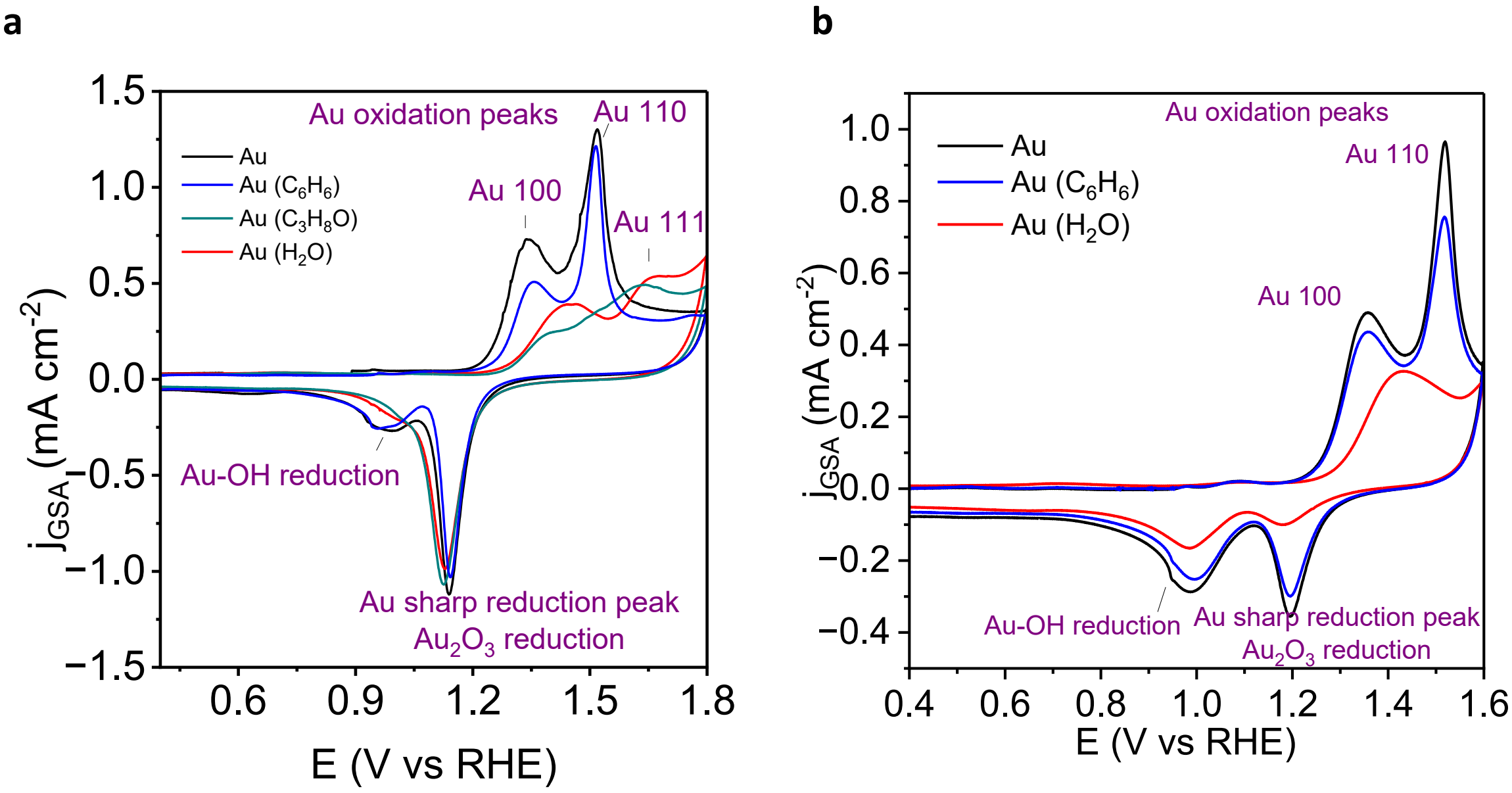

**Figure S16:** (a) Cyclic voltammetry of SeWS Janus before and after treatment in various solvents. A very clear 100 and 110 peak is available for pure Au, whereas it diminishes for the enhancement of the 111 peak in case of $H_2O$ or IPA sonicated samples. (b) CV data scanned till 1.6V to avoid the 111 peak. This diminishes the sharp $Au_2O_3$ reduction peak for $H_2O$ sonicated samples, whereas the highest reduction peak remains intact for Bare Au as the $Au_2O_3$ peak.

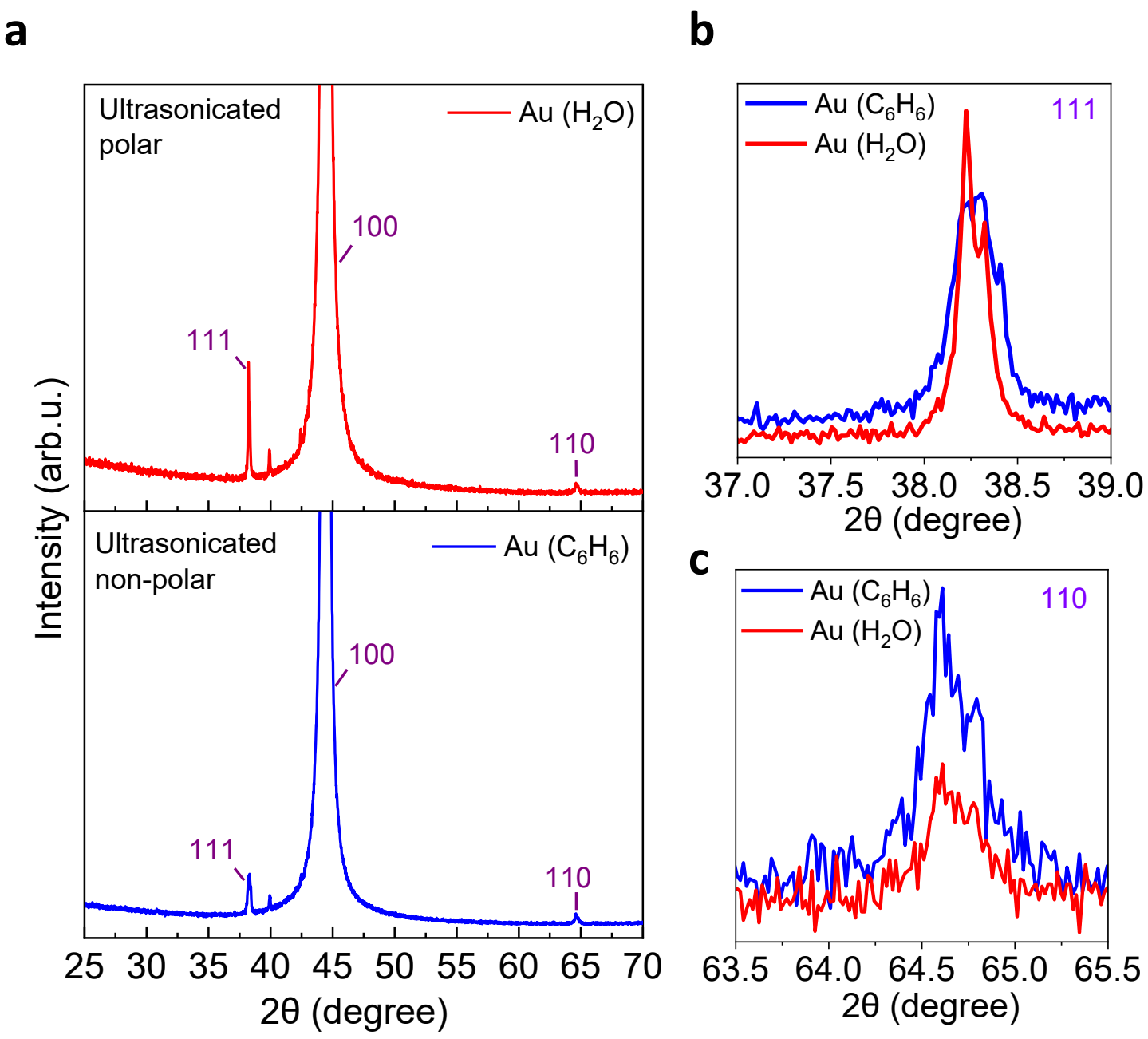

**Figure S17:** (a) Thin film XRD crystallographic data for Au ($H_2O$) and Au ($C_6H_6$) showing all three characteristic peaks of the (111), (100), (110) surfaces (ICDD reference number 00-004-0784). (b) and (c) show the intensity difference between the (111) and (110) surfaces.

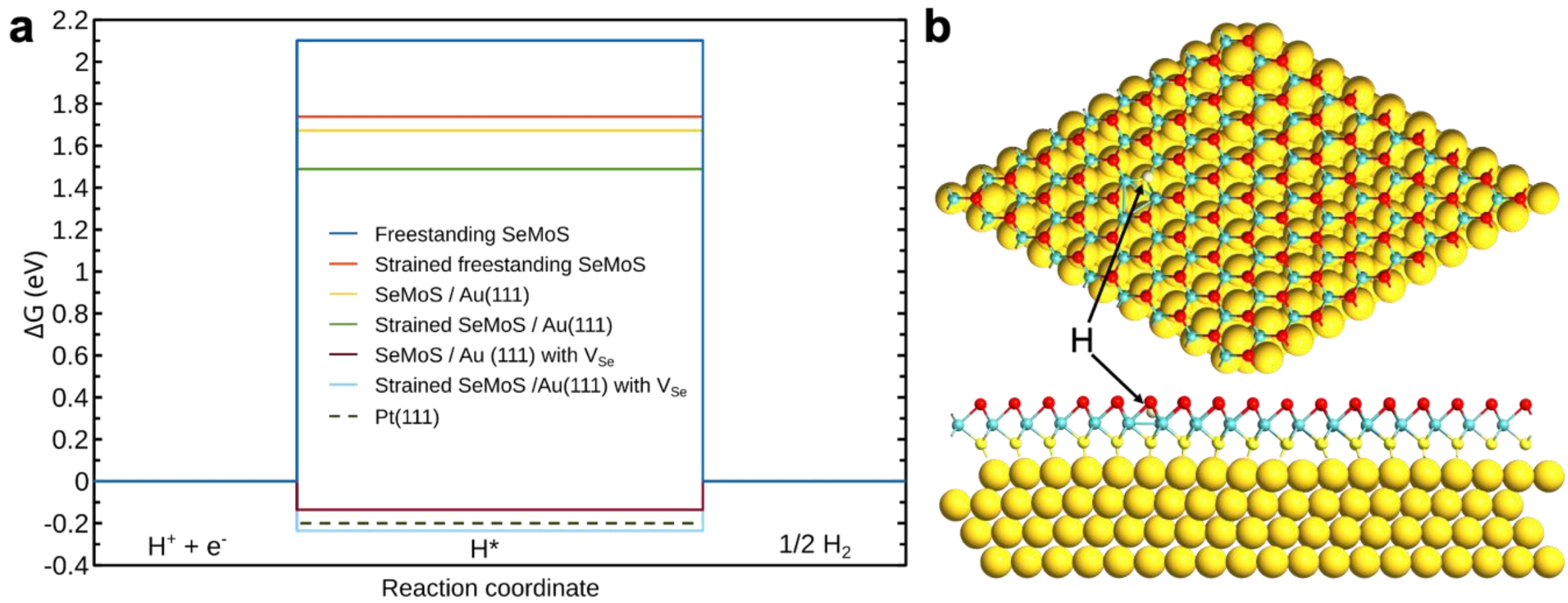

**Figure S18:** (a) Gibbs free energy diagram for HER activity on freestanding and supported Janus SeMoS with the effect of Se vacancy and tensile strain. The applied strain corresponds to 6% biaxial tensile strain. (b) The optimised atomic structure of the SeMoS /Au(111) with hydrogen adsorbed on a Se vacancy.

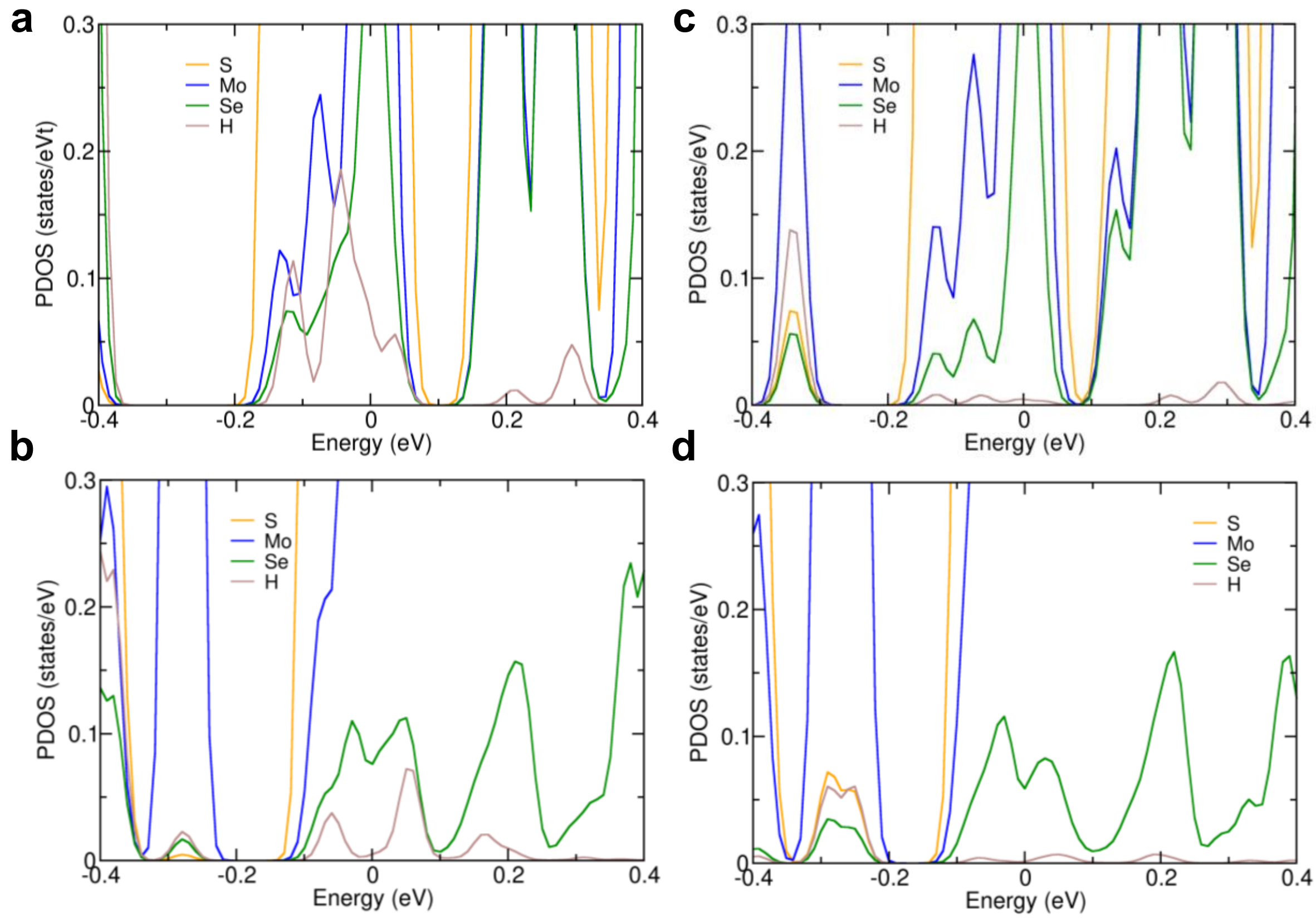

**Figure S19.** Projected density of states (PDOS) for hydrogen adsorption on the Janus SeMoS monolayer on Au(111) substrate: (a) pristine structure, (b) strained structure, (c) Se-vacancy configuration, and (d) strained Se-vacancy structure. The applied strain corresponds to 6% biaxial tensile strain. The Fermi level is set to zero energy.

**Table S1:** Dipole moment of various solvents.

| Solvent | Dipole moment (Debye) |
|---|---|
| Benzene ($C_6H_6$) | 0 |
| Chloroform ($CHCl_3$) | 1.15 |
| IPA ($C_3H_8O$) | 1.66 |
| Water ($H_2O$) | 2.8 |

**Table S2:** Gibbs free energy difference ($\Delta G_H$) for HER activity on freestanding and supported Janus SeMoS with the effect of Se vacancy and tensile strain.

| Catalyst | $\Delta G_H$ (eV) |
|---|---|
| freestanding | 2.10 |
| strained freestanding SeMoS | 1.74 |
| SeMoS /Au(111) | 1.67 |
| strained SeMoS /Au(111) | 1.49 |
| SeMoS /Au(111) with $V_{Se}$ | -0.14 |
| strained SeMoS /Au(111) with $V_{Se}$ | -0.24 |
| Pt(111) | -0.20 |

**References**

[1] A. George, C. Neumann, D. Kaiser, R. Mupparapu, T. Lehnert, U. Hübner, Z. Tang, A. Winter, U. Kaiser, I. Staude, A. Turchanin, *JPhys Mater.* **2019**, *2*, 016001.

[2] Z. Gan, I. Paradisanos, A. Estrada-Real, J. Picker, E. Najafidehaghani, F. Davies, C. Neumann, C. Robert, P. Wiecha, K. Watanabe, T. Taniguchi, X. Marie, J. Biskupek, M. Mundszinger, R. Leiter, U. Kaiser, A. V. Krasheninnikov, B. Urbaszek, A. George, A. Turchanin, *Adv. Mater.* **2022**, *34*, 2205226.

[3] G. Kresse, J. Furthmüller, *Comput. Mater. Sci.* **1996**, *6*, 15.

[4] G. Kresse, J. Furthmüller, *Phys. Rev. B.* **1996**, *54*, 11169.

[5] J. P. Perdew, K. Burke, M. Ernzerhof, *Phys. Rev. Lett.* **1996**, *77*, 3865.

[6] S. Grimme, S. Ehrlich, L. Goerigk, *J. Comput. Chem.* **2011**, *32*, 1456.

[7] S. Pakhira, S. N. Upadhyay, *Sustain. Energy Fuels* **2022**, *6*, 1733.

[8] K. Rahimi, *Int. J. Hydrog. Energy* **2025**, *161*, 150744.

[9] J. K. Nørskov, J. Rossmeisl, A. Logadottir, L. Lindqvist, J. R. Kitchin, T. Bligaard, H. Jónsson, *J. Phys. Chem. B* **2004**, *108*, 17886.